\documentclass[preprint,10pt]{elsarticle}
\biboptions{sort&compress}

\usepackage[a4paper,left=2.0cm,right=2.0cm,top=2.5cm,bottom=2.5cm,footskip=1.6cm]{geometry}
\usepackage{amsmath,amssymb,amsfonts}
\usepackage[scr=boondoxo]{mathalpha}
\let\mathpzc\mathscr
\usepackage{graphicx}
\usepackage{placeins}
\usepackage{booktabs}
\usepackage{algorithm}
\usepackage{algpseudocode}
\usepackage{bm}
\usepackage{xcolor}
\usepackage[hidelinks]{hyperref}
\usepackage[compact]{titlesec}
\usepackage[normalem]{ulem}
\usepackage{xcolor}
\definecolor{hk}{rgb}{0,0,1}
\usepackage{enumitem}
\newsavebox{\strkbox}

\graphicspath{{figures/}}

\titlespacing*{\section}{0pt}{1.0ex plus 0.2ex minus 0.2ex}{0.6ex plus 0.1ex}
\titlespacing*{\subsection}{0pt}{0.8ex plus 0.2ex minus 0.2ex}{0.4ex plus 0.1ex}
\titlespacing*{\subsubsection}{0pt}{0.6ex plus 0.2ex minus 0.2ex}{0.3ex plus 0.1ex}
\makeatletter
\def\ps@pprintTitle{%
  \let\@oddhead\@empty
  \let\@evenhead\@empty
  \def\@oddfoot{\reset@font\hfil\thepage\hfil}%
  \let\@evenfoot\@oddfoot}
\makeatother

\newtheorem{remarkinner}{Remark}
\newenvironment{remark}
  {\begin{remarkinner}\normalfont}
  {\end{remarkinner}}

\begin{document}

\begin{frontmatter}

\title{Feature-preserving Latent-EnKF \\ for Data Assimilation of Flows with Shocks}

\author[1]{Hemanth Chandravamsi}
\author[1]{Hangchuan Hu}
\author[1]{Ponkrshnan Thiagarajan}
\author[1]{Tamer\,\,A.\,\,Zaki\corref{cor1}}
\cortext[cor1]{Corresponding author: \texttt{t.zaki@jhu.edu}}
\address[1]{Department of Mechanical Engineering, Johns Hopkins University, Baltimore, MD 21218, USA}

\begin{abstract}


The ensemble Kalman filter (EnKF) is widely adopted for sequential data assimilation, but fails for solutions with discontinuities, such as shocks in compressible flows. For ensembles with uncertainty in discontinuity location, the EnKF analysis can lose physical consistency, which can be viewed through three lenses: (i) bimodal pointwise statistics in the forecast ensemble, (ii) mixed-sign analysis weights in the affine superposition of forecast members, and (iii) spatially distributed Kalman-gain profiles that modify the state around a discontinuity rather than translating it coherently. In order to preserve flow features in the analysis state and, as such, maintain physical consistency, a feature-preserving latent-EnKF is introduced. The proposed method performs the ensemble update in a low-dimensional latent space learned by a coordinate-conditioned auto-decoder. The latent representation retains near-Gaussian ensemble statistics and provides coordinates in which flow features vary coherently across the forecast ensemble. The EnKF updates performed on the latent codes are decoded into coherent changes in the represented flow features, translating shocks through coordinate morphing rather than superposing them at displaced locations. The corresponding Kalman-gain profiles generated by the latent-EnKF exhibit an approximately step-like variation across the discontinuities, allowing the fronts to translate while preserving sharp features and monotonicity. Sod shock-tube, Mach~2 shock-cylinder, and immersed-body shape-inference tests with sparse, noisy observations demonstrate accurate recovery of shocks and contact discontinuities without spurious oscillations.

\end{abstract}


\end{frontmatter}

{\color{blue}
\renewcommand{\contentsname}{Table of contents \textcolor{red}{[not included in the final paper]}}
\setcounter{tocdepth}{3}
}

\section{Introduction}
\label{sec:intro}

Sequential data assimilation (DA) combines a forecast state, obtained by advancing the flow equations numerically, with noisy observations obtained from the ground truth to produce a posterior estimate of the flow state \cite{kalnay2003,carrassi2018}. The ensemble Kalman filter (EnKF) \cite{evensen2009data,evensen2022fundamentals} is the most widely used sequential DA technique, in which the state is inferred through a sequence of `forecast' and `analysis' steps \cite{zaki2025data}. In the forecast step, the numerical model advances an ensemble of possible flow states in time, with the spread of the ensemble representing the uncertainty in the prediction. When observations become available, the analysis step uses them to correct the forecast. The correction is determined by how strongly different parts of the flow vary with the observed quantities across the ensemble, allowing information from a limited number of measurements to update portions of the flow that are not directly observed. Repeating these forecast and analysis steps as new measurements become available provides a repeatedly updated estimate of the flow state. In practice, flow measurements are often sparse and noisy and may provide only indirect information about the underlying state. In high-speed flows, these measurements may include schlieren images, wall-pressure probes, and velocimetry. Assimilation of these measurements with a numerical forecast through the EnKF can provide an estimate of the full flow state. However, the presence of shocks introduces an additional and distinct challenge. Discontinuous features in particular, such as shocks, material interfaces, and boundaries, can limit the accuracy of EnKF data assimilation, as established in previous works spanning both variational and sequential formulations \cite{houba2024,hansen2024,edoh2025,zhou2026neural,west2025variational,edoh2026}. However, the mechanisms responsible for this loss of accuracy are not yet fully understood. The present work builds on previous studies of data assimilation and state estimation in compressible flows \cite{buchta2021observation,buchta2022assimilation,wang2025domain,morra2024ml,zhang2022ensemble} and in flows containing shocks \cite{ke2026observability}, with particular focus on understanding and addressing this difficulty. 

Although the EnKF is an established and efficient method for sequential data assimilation in flows with smooth features \cite{zaki2025data,zaki2025turbulence,asch2016}, the analysis state can become unreliable when the forecast contains discontinuities with uncertain locations. The EnKF represents forecast uncertainty through the ensemble mean and covariance, an approximation that is most appropriate when the forecast statistics are approximately Gaussian \cite{evensen2009data,burgers1998}. This representation becomes inadequate near a discontinuity whose location varies across the ensemble. In shock-laden flows, uncertainty in shock position means that, at a given spatial location, some ensemble members may lie upstream of the shock while others lie downstream. This leads to multimodal state distribution that cannot be adequately represented by a single mean and covariance \cite{houba2024,morzfeld2019,zhou2026neural}.

A related difficulty arises from the linear nature of the EnKF analysis update, which constrains each analysis member to the affine subspace spanned by the forecast ensemble \cite{iglesias2013}. repeatedly analysis combines forecast states through linear superposition in the physical state space. When ensemble variability is dominated by uncertainty in discontinuity location, such combinations do not, in general, correspond to a coherent displacement of the discontinuity. Instead, such combinations can produce non-physical intermediate states containing spurious oscillations or multiple discontinuities. This behavior can be understood geometrically: if the ensemble consists of states containing the same discontinuity at different locations, a pointwise linear combination of these states generally produces multiple jumps, whereas a physically admissible state contains a single discontinuity at an updated location.

Several approaches have been proposed to alleviate the difficulties posed by discontinuities, including positivity-preserving transformations \cite{edoh2025}, rank-based normal-score transformations \cite{hansen2024,zhou2011}, and structurally informed covariance models \cite{li2024,li2025structurally,glaubitz2026regularity}. Hansen et al.\ \cite{hansen2025detonation} also aligned ensemble members with respect to the detonation front before applying a local ensemble transform filter, thereby reducing variability associated with front-position uncertainty. While these approaches can improve the analysis, they do not directly eliminate the difficulty associated with combining multiple discontinuities located at different positions. More broadly, non-Gaussian forecast statistics have been addressed using particle filters \cite{snyder2008obstacles,poterjoy2016localized}, Gaussian mixture filters \cite{bengtsson2003toward,hoteit2008new,stordal2011adaptive,smith2007cluster,sondergaard2013parti,sondergaard2013partii}, rank-histogram filters \cite{anderson2010non}, and anamorphosis filters \cite{grooms2022comparison,bertino2003sequential,simon2009application}. Related treatments include iterative ensemble smoothers \cite{bocquet2014iterative}, measure-transport and score-based methods \cite{spantini2022coupling,ramgraber2023ensemble,rozet2023score}, and generative filtering that samples a Gaussian mixture fitted to the forecast ensemble \cite{zhou2026ensemble}. These methods modify how the forecast distribution is represented or updated, but the resulting analysis is still formed from pointwise combinations of states in the original physical space. A smaller body of work instead seeks to move coherent features to new locations through field alignment \cite{ravela2007data}, image morphing \cite{beezley2008morphing}, displacement corrections \cite{nehrkorn2015correcting}, or optimal transport \cite{feyeux2018optimal,tamang2021ensemble,subrahmanya2025}. Such methods can account explicitly for positional uncertainty, but typically require the relevant features to be identified in each member and a displacement field or transport map to be constructed or prescribed.

Instead of performing the EnKF update directly on the physical state, Zhou et al.\ \cite{zhou2026neural} perform the analysis in the parameter space of deep neural networks used to represent individual ensemble members. This neural EnKF is effective, but requires sequential network training along a nearest-neighbor chain to maintain coherence among the ensemble members in weight space. A related class of methods instead transforms the physical state into a learned latent space \cite{fan2026physically,chen2023reduced}. Tong et al.\ \cite{tong2026latent}, for example, perform the EnKF analysis in the latent space of an autoencoder coupled with learned linear latent dynamics for strongly nonlinear systems. More generally, latent-space assimilation methods encode forecast members into low-dimensional representations, update the corresponding latent variables, and then decode them back to the physical state \cite{amendola2021data,peyron2021latent,cheng2023machine,fan2025novel,mousavi2025}. Variational autoencoders have also been used to obtain latent statistics that are closer to Gaussian than those of the physical state \cite{pasmans2026ensemble}, while neural operators and learned filters have been trained to perform the analysis directly \cite{zhou2024bieqno,bach2025}. With the exception of the neural EnKF, these approaches rely on an encoder trained offline on a data set that must adequately represent the states encountered during online assimilation, where out-of-distribution states may otherwise occur.

The present work shows that neither sequential chain training nor a separately trained latent encoder is required. Instead, the EnKF analysis is performed in the latent space of a coordinate-conditioned auto-decoder (CC-auto-decoder) \cite{park2019deepsdf}, which provides a continuous and fully differentiable representation of the flow field. For each ensemble member, the decoder maps a latent code together with spatial coordinates to the corresponding flow variables. The decoder weights and member-specific latent codes are optimized jointly on the forecast ensemble, placing the ensemble on a smooth, low-dimensional manifold parameterized by the latent variables. Within this representation, sharp physical-space features, including shocks, contact discontinuities, and slip lines, can be represented through smooth variations in latent space. The adopted CC-auto-decoder formulation also reduces the amount of training data required relative to conventional encoder-decoder approaches for learning accurate low-dimensional encoding of the sharp features. The key contributions of this work are:
\begin{itemize}\setlength{\itemsep}{1pt}\setlength{\parskip}{0pt}

\item[(i)] A physical and statistical explanation of why the EnKF fails near discontinuities (Sec.~\ref{sec:failure}).

\item[(ii)] A feature-preserving latent-EnKF that performs the analysis in a learned low-dimensional latent space, demonstrated through data assimilation experiments with sparse and noisy observations to reconstruct the full state of shock-laden flows and extended to infer the geometry of an unknown immersed body from surrounding flow measurements (Secs.~\ref{sec:method}, \ref{sec:experiments}, and \ref{sec:cylgeom}).

\item[(iii)] An explanation of how the latent codes and shared decoder represent the ensemble states, and why this representation is well suited to the EnKF analysis (Secs.~\ref{sec:mechanism} and \ref{sec:representation}).

\end{itemize}

The paper is organized as follows. Section~\ref{sec:failure} examines why the physical-space EnKF fails near discontinuities, while Section~\ref{sec:method} introduces the feature-preserving latent EnKF. The assimilation experiments are presented in Section~\ref{sec:experiments}. Section~\ref{sec:mechanism} explains how the latent-space analysis resolves the failure of the physical-space update, and Section~\ref{sec:representation} examines how the CC-auto-decoder represents the ensemble through coordinate morphing and shared basis functions. Section~\ref{sec:cylgeom} extends the latent EnKF to infer the geometry of an immersed body from flow observations, and Section~\ref{sec:conclusions} summarizes the main conclusions.

\section{Physical-space EnKF and the origin of oscillations near discontinuities}
\label{sec:failure}

\begin{figure}[!t]
\centering
\includegraphics[width=0.95\textwidth]{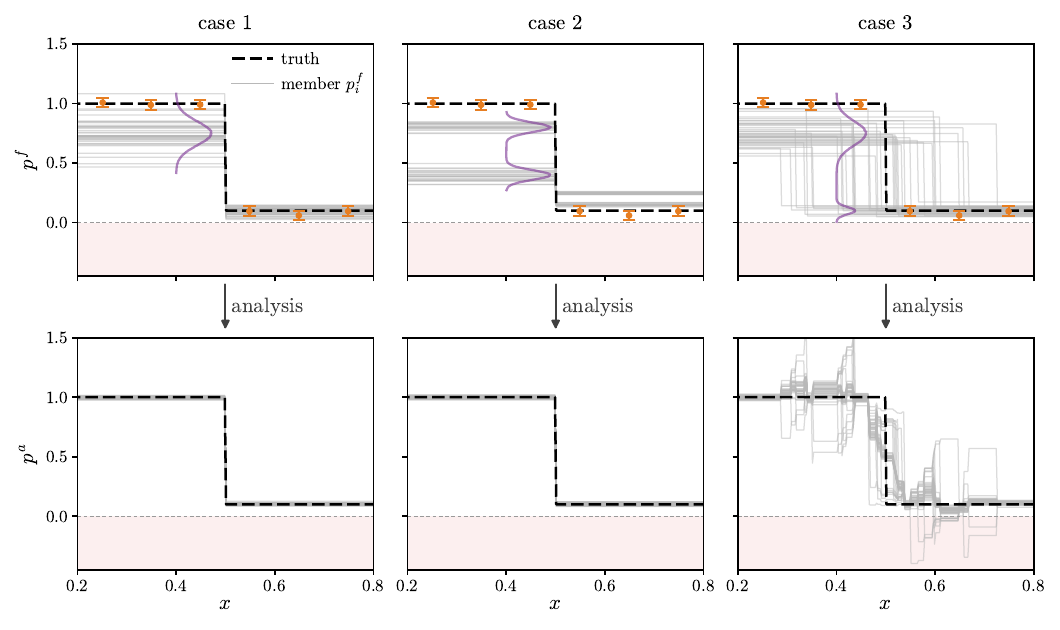}
\caption{Physical-space EnKF analysis of three step ensembles with the same truth and ten observations; $n_e=40$. Columns, left to right: case 1, aligned fronts; case 2, aligned fronts with two clusters of states on each side; case 3, misaligned fronts. Top row: forecast; bottom row: analysis. The sideways curve is the pointwise probability density of $p$ at $x=0.40$.}
\label{fig:physical-enkf-cases}
\end{figure}

This section introduces the EnKF in physical space and explains the mechanism by which oscillations arise near a discontinuity. Let $\mathbf{q}\in\mathbb{R}^{n_x}$ denote the discretized flow state on the spatial grid $\mathbf{x}$. The forecast ensemble is denoted by $\{\mathbf{q}_i^{f}\}_{i=1}^{n_e}$. Each forecast member is obtained by advancing the corresponding previous analysis member, $\mathbf{q}_i^{a}$, over one observation interval $\Delta t$ with the forecast model $\mathcal{F}_{\Delta t}$. For the first cycle, the corresponding member of the initial ensemble at $t=0$ serves as the initial guess. The observation operator $\mathcal{H}:\mathbb{R}^{n_x}\rightarrow\mathbb{R}^{n_y}$ then maps each forecast state to the corresponding predicted sensor measurements,
\begin{equation}
\mathbf{q}_i^{f}=\mathcal{F}_{\Delta t}(\mathbf{q}_i^{a}),\qquad
\mathbf{y}_i^{f}=\mathcal{H}(\mathbf{q}_i^{f}),\qquad i=1,\dots,n_e.
\end{equation}
The observations are perturbed independently for each ensemble member as $\mathbf{d}+\bm{\eta}_i$, where $\bm{\eta}_i\sim\mathcal{N}(\mathbf{0},\mathbf{R})$. Here, $\mathbf{R}$ is the observation-error covariance matrix and is taken to be diagonal, so the observation noise at different sensors is uncorrelated. The forecast ensemble means are defined as
\begin{equation}
\bar{\mathbf{q}}^{f}=\frac{1}{n_e}\sum_{i=1}^{n_e}\mathbf{q}_i^{f},
\qquad
\bar{\mathbf{y}}^{f}=\frac{1}{n_e}\sum_{i=1}^{n_e}\mathbf{y}_i^{f},
\end{equation}
with the corresponding normalized perturbation matrices
\begin{equation}
\label{eq:perturb}
\mathbf{Q}=
\frac{1}{\sqrt{n_e-1}}
\bigl[\mathbf{q}_i^{f}-\bar{\mathbf{q}}^{f}\bigr]_i
\in\mathbb{R}^{n_x\times n_e},
\qquad
\mathbf{Y}=
\frac{1}{\sqrt{n_e-1}}
\bigl[\mathbf{y}_i^{f}-\bar{\mathbf{y}}^{f}\bigr]_i
\in\mathbb{R}^{n_y\times n_e}.
\end{equation}
Each forecast member of $\mathbf{q}_i^{f}$ is then corrected using the discrepancy between its predicted observations and the corresponding perturbed observations through the EnKF update
\begin{equation}
\label{eq:enkf-physical}
\mathbf{q}_i^{a}
=
\mathbf{q}_i^{f}
+
\underbrace{
\mathbf{Q}\mathbf{Y}^{\!\top}
\bigl(
\mathbf{Y}\mathbf{Y}^{\!\top}+\mathbf{R}
\bigr)^{-1}
}_{\text{Kalman gain }\mathbf{K}}
\underbrace{
\bigl(
\mathbf{d}+\bm{\eta}_i-\mathbf{y}_i^{f}
\bigr)
}_{\text{observation mismatch}}.
\end{equation}
Here, $\bm{\eta}_i$ represents the observation noise associated with ensemble member $i$. The linear EnKF update in Equation~\eqref{eq:enkf-physical} is examined throughout the remainder of this section to identify the conditions under which oscillations develop near a discontinuity.

To diagnose the generation of oscillations near a discontinuity, a one-dimensional test problem consisting of a scalar field with a single jump is considered on $x\in[0,1]$. Three EnKF experiments are used to assess the failure of the physical-space EnKF near the discontinuity. All three experiments use the same truth, with a discontinuity at $x_s=0.5$, $p=1.0$ for $x<x_s$, and $p=0.1$ for $x>x_s$. Ten equally spaced sensors sample the truth with observation noise of standard deviation $0.02$. Each initial-guess ensemble contains $n_e=40$ members, with member $i$ defined by $p_i(x)=p_L^{(i)}$ for $x<x_s^{(i)}$ and $p_R^{(i)}$ elsewhere. The experiments differ in the distributions of $p_L^{(i)}$, $p_R^{(i)}$, and $x_s^{(i)}$, giving the following three cases, shown from left to right in Fig.~\ref{fig:physical-enkf-cases}:
\begin{enumerate}
\item \emph{Case 1}, aligned fronts, $x_s^{(i)}=0.5$, $p_L^{(i)}\sim\mathcal{N}(0.75,0.10^2)$, $p_R^{(i)}\sim\mathcal{N}(0.1,0.03^2)$;
\item \emph{Case 2}, aligned fronts with bimodal states on each side: left of the front, $\mathcal{N}(0.8,0.04^2)$ and $\mathcal{N}(0.4,0.04^2)$ with equal probability; right of the front, $\mathcal{N}(0.15,0.01^2)$ and $\mathcal{N}(0.25,0.01^2)$; the pointwise probability density then has two peaks at every $x$;
\item \emph{Case 3}, misaligned fronts, $x_s^{(i)}\sim\mathcal{N}(0.5,0.10^2)$, with the states of case 1.
\end{enumerate}

A single analysis update is applied to each ensemble using Eqn.~\eqref{eq:enkf-physical}. Figure~\ref{fig:physical-enkf-cases} compares the forecast and analysis ensembles for the three cases. When the front is aligned across the ensemble, as in cases 1 and 2, the EnKF preserves the single-step profile and monotonicity. This behavior remains true with case 2 ensemble despite its bimodal pointwise probability distribution, indicated by the sideways curve at $x=0.40$. The outcome changes markedly in case 3, where the front position varies across members. Analysis on Case 3 ensemble develops additional fronts, substantially larger overshoots, and negative pressures in several members. The contrast between cases 2 and 3 shows that \emph{bimodal pointwise statistics alone are not sufficient to generate oscillations}. The results of figure~\ref{fig:physical-enkf-cases} shows that the contrast in the analysis state rests primarily on the front alignment. The necessary conditions under which such front misalignment and the role of pointwise bimodality leads to a non-monotone analysis states are discussed in this section. In what follows, the failure mechanism associated with the misaligned-fronts is examined. 



\subsection{The generation of bimodal statistics in ensembles of discontinuous functions}
\label{sec:bimodal}

In this section, we derive the pointwise probability distribution of an ensemble of step functions and show how uncertainty in front location gives rise to bimodal statistics near the discontinuity. We consider member-dependent plateau states and advection speeds, with the step of member $i$ advecting according to,
\begin{equation*}
p_i(x,t)=
\begin{cases}
p_L^{(i)}, & x<x_s+c_it,\\
p_R^{(i)}, & x>x_s+c_it,
\end{cases}
\end{equation*}
with independently sampled plateau states and advection speeds,
\begin{equation*}
p_L^{(i)}\sim\mathcal{N}(\mu_L,\sigma_L^2),\qquad
p_R^{(i)}\sim\mathcal{N}(\mu_R,\sigma_R^2),\qquad
c_i\sim\mathcal{N}(0,\sigma_c^2).
\end{equation*}
At $t=0$, all ensemble members are initialized with the same front location $x=x_s$. At a later time $t>0$, the front of member $i$ is located at $x_s+c_it$. Hence, at a given location $x$, members whose fronts lie to the right of $x$ contribute values from the left plateau $p_L$, while members whose fronts lie to the left contribute values from the right plateau $p_R$.

\begin{figure}[!t]
\centering
\includegraphics[width=0.95\textwidth]{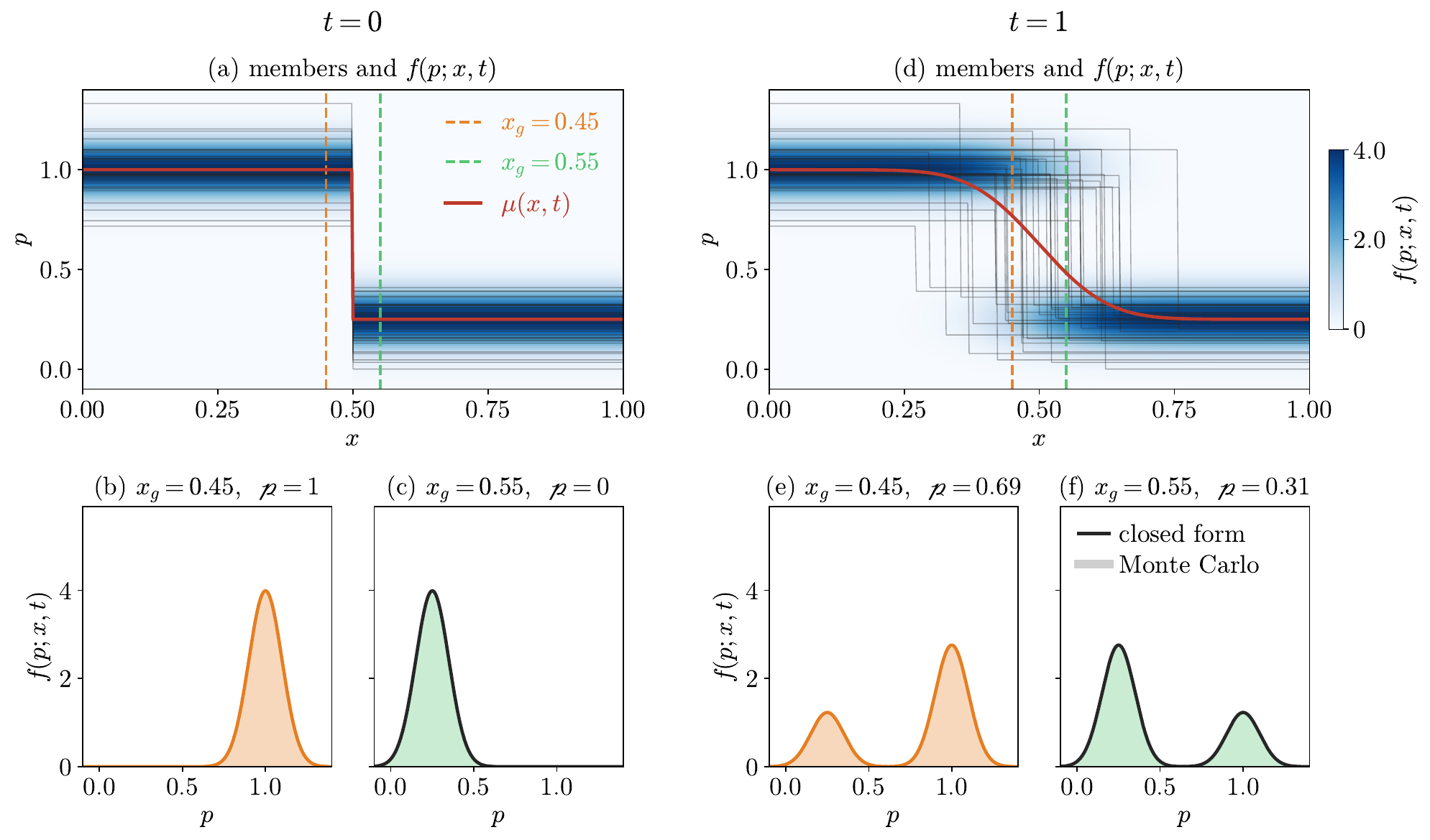}
\caption{\textbf{Analytical bimodal statistics of an ensemble of shocks with misaligned positions.} Ensemble members over the closed-form probability density $f(p;x,t)$ from Eqn.~\eqref{eq:mixture-density}, together with the probability density at generic locations $x_g=0.45$ and $x_g=0.55$ compared with a Monte Carlo ensemble. The left block shows $t=0$ and the right block $t=1$. The value of $\mathpzc{p}$ shown at each station is the probability that a member samples the state to the left of the shock.}
\label{fig:step-bimodal}
\end{figure}

At a fixed $(x,t)$, define
\begin{equation}
    \mathpzc{p}(x,t) = \mathbb{P}(x_s+c_it>x),  
\end{equation}
the probability that the front of a randomly selected ensemble member lies to the right of $x$. A member takes its left-state value when this event occurs and its right-state value otherwise. Rearranging the terms of the event $x_s+c_it>x$ gives
\[
\mathpzc{p}(x,t)
=
\mathbb{P}\!\left(
c_i>\frac{x-x_s}{t}
\right)
=
\mathbb{P}\!\left(
\frac{c_i}{\sigma_c}>
\frac{x-x_s}{\sigma_c t}
\right).
\]
Since $c_i/\sigma_c$ is standard normal,
\[
\mathpzc{p}(x,t)
=
1-\Phi\!\left(\frac{x-x_s}{\sigma_c t}\right)
=
\Phi\!\left(\frac{x_s-x}{\sigma_c t}\right).
\]
Here, $\Phi(\xi)=\frac{1}{\sqrt{2\pi}}\int_{-\infty}^{\xi} e^{-s^2/2}\,ds$ is the standard normal cumulative distribution function. Since $c_i$ is independent of $p_L^{(i)}$ and $p_R^{(i)}$, the pointwise pressure distribution is a mixture of the left- and right-state distributions, weighted by the probabilities of sampling each side of the front. Therefore,
\begin{equation}
\label{eq:mixture-density}
f(p;x,t)
=
\mathpzc{p}\,\mathcal{N}(\mu_L,\sigma_L^2)
+
(1-\mathpzc{p})\,\mathcal{N}(\mu_R,\sigma_R^2).
\end{equation}

The corresponding mean of the distribution in Eqn.~\ref{eq:mixture-density} interpolates between $\mu_L$ and $\mu_R$ across the region occupied by the fronts, so the ensemble mean carries a smooth transition while every member retains a sharp jump. At a fixed location $x$, a fraction $\mathpzc{p}$ of the ensemble samples the left plateau and a fraction $1-\mathpzc{p}$ samples the right plateau. The ensemble mean is therefore the weighted contribution,
\begin{equation}
\mu(x,t)
=
\mathpzc{p}\,\mu_L
+
(1-\mathpzc{p})\,\mu_R.
\end{equation}
The pointwise variance contains two contributions. The first comes from the variability within the left and right plateau states,
\begin{equation*}
\mathpzc{p}\,\sigma_L^2
+
(1-\mathpzc{p})\,\sigma_R^2.
\end{equation*}
The second comes from the separation between the two plateau means. Since the ensemble samples $\mu_L$ with probability $\mathpzc{p}$ and $\mu_R$ with probability $1-\mathpzc{p}$, this contribution is
\begin{equation*}
\mathpzc{p}(1-\mathpzc{p})(\mu_L-\mu_R)^2.
\end{equation*}
Adding the two contributions gives
\begin{equation}
\label{eq:mixture-moments}
\sigma^2(x,t)
=
\mathpzc{p}\,\sigma_L^2
+
(1-\mathpzc{p})\,\sigma_R^2
+
\mathpzc{p}(1-\mathpzc{p})(\mu_L-\mu_R)^2.
\end{equation}

Although these analytical results strictly apply in the large-ensemble limit, finite ensembles exhibit the same qualitative behavior. Equation~\eqref{eq:mixture-density} makes the origin of the bimodality explicit. Outside the region occupied by the dispersed fronts, $\mathpzc{p}$ approaches either $0$ or $1$, and the pointwise distribution reduces to one of the two Gaussian plateau distributions. Within the region occupied by the fronts, both plateau distributions contribute. Figure~\ref{fig:step-bimodal} compares the pointwise probability density before and after the fronts disperse from $t=0 \rightarrow 1$, for $\mu_L=1.0$, $\mu_R=0.25$, $\sigma_L=\sigma_R=0.10$, $\sigma_c=0.10$, and $x_s=0.5$. At $t=0$, all fronts coincide at $x_s$, so $x_g=0.45$ samples only the left plateau and $x_g=0.55$ samples only the right plateau. The corresponding distributions are therefore unimodal. At $t=1$, the front locations have standard deviation $\sigma_c t=0.1$. The two stations now lie within the region occupied by the dispersed fronts, with $\mathpzc{p}=0.69$ at $x_g=0.45$ and $\mathpzc{p}=0.31$ at $x_g=0.55$. Both plateau distributions therefore contribute at each station, producing the bimodal distributions in panels~(e) and~(f). The closed-form densities agree closely with the Monte Carlo histograms obtained from $2\times10^{6}$ samples. The derivation therefore shows that front-location uncertainty naturally produces bimodal pointwise statistics near the discontinuity, violating the Gaussian assumption underlying the standard EnKF update. However, as demonstrated by case 2 in Fig.~\ref{fig:physical-enkf-cases}, pointwise bimodality alone is not sufficient to produce non-monotone analysis states. The next two subsections examine the additional mechanisms by which front misalignment leads to oscillations.



\subsection{Origin of non-monotonicity viewed through weighted superposition of ensemble members}
\label{sec:superposition-view}


The EnKF analysis state shown in Eqn.~\eqref{eq:enkf-physical} can be rewritten algebraically as a weighted linear superposition of the forecast ensemble members \cite{evensen2003ensemble,evensen2009data},
\begin{equation}
\label{eq:affine}
\mathbf{q}_i^{a}
=
\sum_{j=1}^{n_e} w_{j}^{(i)}\,\mathbf{q}_j^{f}.
\end{equation}
For analysis member $i$, the weights are $w_{j}^{(i)}=\delta_{ij}+c_{j}^{(i)}$, where $\delta_{ij}$ is the Kronecker delta and
\begin{equation}
\label{eq:weights-closed}
c_{j}^{(i)}
=
\frac{1}{\sqrt{n_e-1}}
\left[
\mathbf{Y}^{\!\top}
\bigl(\mathbf{Y}\mathbf{Y}^{\!\top}+\mathbf{R}\bigr)^{-1}
\bigl(\mathbf{d}+\bm{\eta}_i-\mathbf{y}_i^{f}\bigr)
\right]_j.
\end{equation}
Since $\sum_{j=1}^{n_e} w_j^{(i)}=1$, the linear superposition of Eqn.~\eqref{eq:affine} defines an affine combination of the forecast members. However, the combination is not necessarily convex because the weights are neither constrained to be positive nor bounded to remain within the range $[0,1]$. For a given analysis member $i$, each weight $w_j^{(i)}$ is independent of spatial position and therefore multiplies the entire forecast member $\mathbf{q}_j^{f}$. The sign of $w_j^{(i)}$ depends on the relation between the observation anomaly associated with forecast member $j$ and the observation mismatch of member $i$, so negative weights can arise naturally. 

\begin{remark}
The fact that the EnKF analysis state is an affine combination of the forecast ensemble members, rather than a convex combination, is central to the loss of monotonicity. Consider one-dimensional forecast members that are individually monotone decreasing (like the ensemble members of cases 1, 2, and 3 of Fig.~\ref{fig:physical-enkf-cases}). For any $x_1<x_2$, $p_j^f(x_1)\ge p_j^f(x_2)$ for all $j$. If the affine weights were all positive, then
\[
p_i^a(x_1)-p_i^a(x_2)
=
\sum_{j=1}^{n_e} w_j^{(i)}
\left[p_j^f(x_1)-p_j^f(x_2)\right]
\ge 0,
\]
and the analysis member $p_i^a$ would necessarily remain monotone decreasing. Thus, non-monotonicity cannot arise from front misalignment alone when the weights are positive. At least one negative weight is therefore necessary for the EnKF update to produce a non-monotone analysis state.
\end{remark}

\begin{figure}[!t]
\centering
\includegraphics[width=0.95\textwidth]{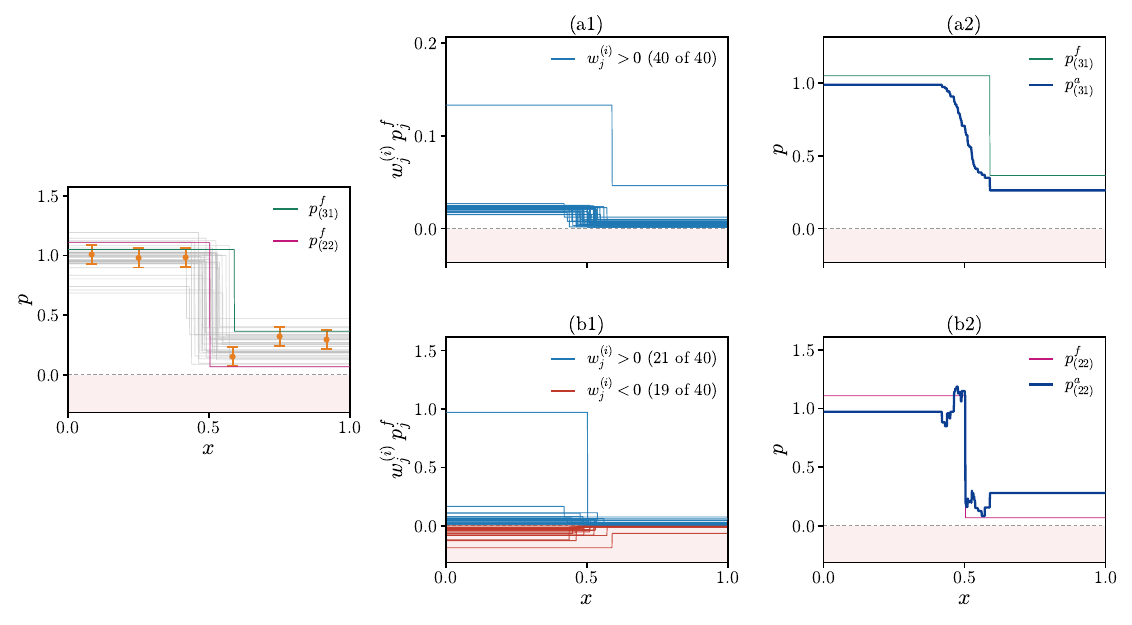}
\caption{The weighted-superposition view shows that mixed-sign weights produce non-monotone analysis states, whereas positive weights preserve monotonicity. The left most panel shows the forecast ensemble, the six observations, and the two selected forecast members, $i=22$ and $i=31$. Panels (a1) and (b1) show the forty weighted forecast states $w_j^{(i)}p_j^f$ that contribute to the analysis states of members $i=22$ and $i=31$, respectively. Panels (a2) and (b2) show the corresponding forecast and analysis states for the $i=22$ and $i=31$ members.}
\label{fig:superposition-view}
\end{figure}


Figure~\ref{fig:superposition-view} illustrates the mechanism identified in Remark 1 using a simple ensemble of forty forecast members. Two members, $i=31$ and $i=22$, are examined in isolation to contrast the effects of positive and mixed-sign weights. For member $i=31$, whose forecast front is located at $x=0.590$, all forty weights are positive, with a minimum value of $0.020$ (figure 3(a1)). The corresponding analysis member therefore remains monotone (figure 3(a2)). In contrast, for member $i=22$, whose forecast front is located at $x=0.503$, nineteen of the forty weights are negative, with a minimum value of $-0.179$. The negative weights reverse the direction of the forecast jumps (see the terms corresponding to $w_{j}^{(i)}<0$ in panel (b1)) and contribute to local increase in the analysis state. Consequently, the analysis member becomes non-monotone and ranges from $0.083$ to $1.186$, exceeding the upper plateau value of $1.0$ as seen in Figure~\ref{fig:superposition-view}(b2). A corresponding undershoot appears near the lower plateau as well, which can in some instances drive thermodynamic variables to non-physical values, such as negative pressure. To conclude, the weighted-superposition view presented in this section identifies the immediate origin of the non-monotonicity: the EnKF combines spatially displaced discontinuities using affine weights of mixed sign.

\subsection{Origin of non-monotonicity viewed through the Kalman gain profiles}
\label{sec:gain}

The non-monotonicity in case 3 ensemble can also be understood from a complementary perspective by examining the spatial profiles of the Kalman corrections associated with measurements from individual sensors. Writing the gain matrix in Eqn.~\eqref{eq:enkf-physical} as
\begin{equation}
\label{eq:gain}
\mathbf{K}
=
\mathbf{Q}\mathbf{Y}^{\!\top}
\bigl(\mathbf{Y}\mathbf{Y}^{\!\top}+\mathbf{R}\bigr)^{-1}
\in\mathbb{R}^{n_x\times n_y},
\end{equation}
the correction to analysis member $i$ can be decomposed into contributions from the individual sensors,
\begin{equation}
\label{eq:gain-columns}
\mathbf{q}_i^{a}-\mathbf{q}_i^{f}
=
\mathbf{K}
\bigl(\mathbf{d}+\bm{\eta}_i-\mathbf{y}_i^{f}\bigr)
=
\sum_{\ell=1}^{n_y}
\mathbf{k}_{\ell}
\underbrace{
\bigl[\mathbf{d}+\bm{\eta}_i-\mathbf{y}_i^{f}\bigr]_{\ell}
}_{\substack{\text{observation mismatch}\\ \text{at sensor }\ell}},
\qquad
\mathbf{K}=
\underbrace{
\bigl[\mathbf{k}_1,\dots,\mathbf{k}_{n_y}\bigr]
}_{\substack{\text{Kalman gain kernels}\\ \text{for individual sensors}}},
\qquad
\mathbf{k}_{\ell}\in\mathbb{R}^{n_x}.
\end{equation}
Each sensor therefore contributes a spatial gain profile $\mathbf{k}_{\ell}$, which is multiplied by the scalar observation mismatch at that sensor to produce its contribution to $\mathbf{q}_i^{a}-\mathbf{q}_i^{f}$. The stronger the correlation between the state at sensor $\ell$ and the state at a spatial location $x$, the larger the corresponding gain value $k_{\ell}(x)$. A zero correlation implies no influence from sensor $\ell$ on the analysis state at $x$, i.e., $k_{\ell}(x)=0$. This relation can be interpreted directly through a linear regression of the ensemble states against the predicted sensor measurements. At a given grid point $x$, $k_{\ell}(x)$ is the slope of the best-fit line relating the forecast state $p^{f}(x)$ to the predicted measurement at sensor $\ell$, as derived in~\ref{app:gain-regression}. For a single-sensor update, the fitted regression line can be written as
\begin{equation}
p(x)
=
k_{\ell}(x)\,\widetilde y(x_{\ell})
+
b_{\ell}(x).
\end{equation}
where $\widetilde y_{\ell}^{f}=y_{\ell}^{f}+\eta_{\ell}$ denotes the noisy sensor measurement. For notational simplicity, the ensemble-member index is suppressed here, so that $y_{\ell}^{f}$ and $\eta_{\ell}$ represent $\bigl[\mathbf{y}_i^{f}\bigr]_{\ell}$ and $\bigl[\bm{\eta}_i\bigr]_{\ell}$, respectively. The slope and intercept are
\begin{equation}
k_{\ell}(x)
=
\frac{
\displaystyle
\sum_{i=1}^{n_e}
\left(p_i^{f}(x)-\overline{p^{f}(x)}\right)
\left(\widetilde y_{i\ell}^{f}-\overline{\widetilde y_{\ell}^{f}}\right)
}{
\displaystyle
\sum_{i=1}^{n_e}
\left(\widetilde y_{i\ell}^{f}-\overline{\widetilde y_{\ell}^{f}}\right)^2
},
\qquad
b_{\ell}(x)
=
\overline{p^{f}(x)}
-
k_{\ell}(x)\,\overline{\widetilde y_{\ell}^{f}}.
\end{equation}
Thus, $k_{\ell}(x)$ is the slope of the best-fit line through the ensemble cloud in the sensor-state plane. Evaluating this slope at each grid point gives the spatial gain profile $\mathbf{k}_{\ell}$, which determines how the observation mismatch at sensor $\ell$ is distributed over the state.

\begin{figure}[!t]
\centering
\includegraphics[width=1.0\textwidth]{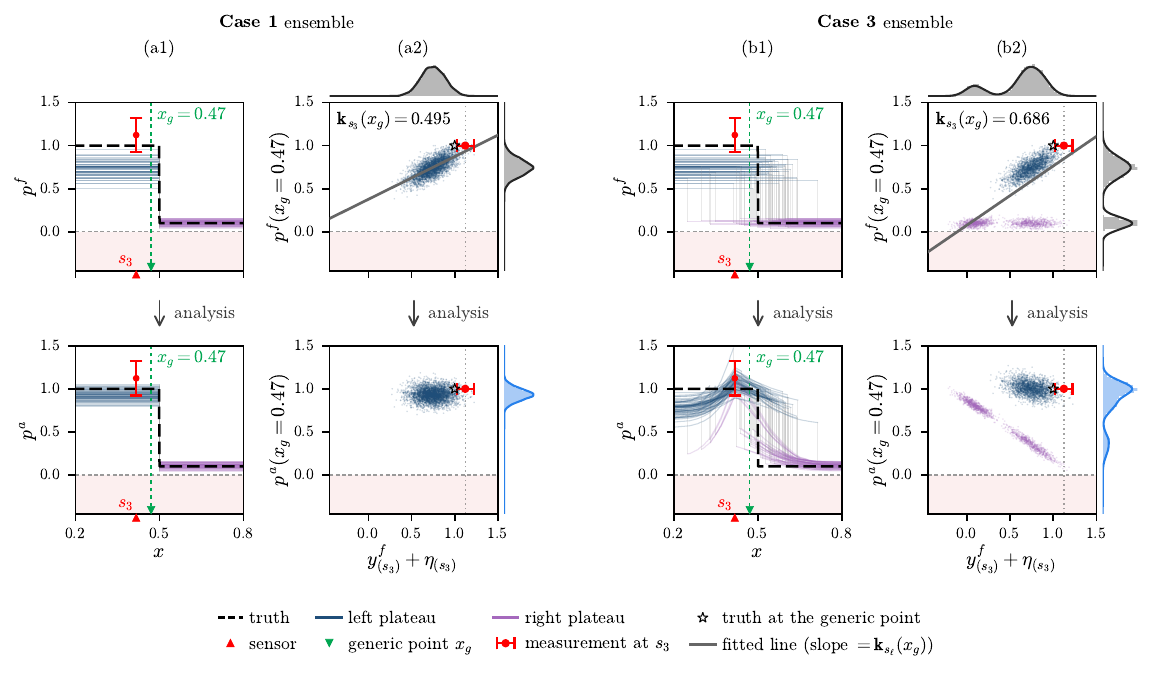}
\caption{Kalman-gain interpretation for cases~1 and~3 using sensor $s_3$ at $x_{(s_3)}=0.417$ and a generic point at $x_g=0.47$. Panels~(a1,b1) show the forecast and analysis ensembles, while panels~(a2,b2) show the corresponding sensor-state distributions and fitted slopes $k_{s_3}(x)$. Case~1 follows a single linear relation, whereas in case~3 the dispersed fronts create distinct local groups that are updated by the same fitted slope, producing the non-monotone analysis response.}
\label{fig:gain-scatter}
\end{figure}

\begin{figure}[!t]
\centering
\includegraphics[width=0.95\textwidth]{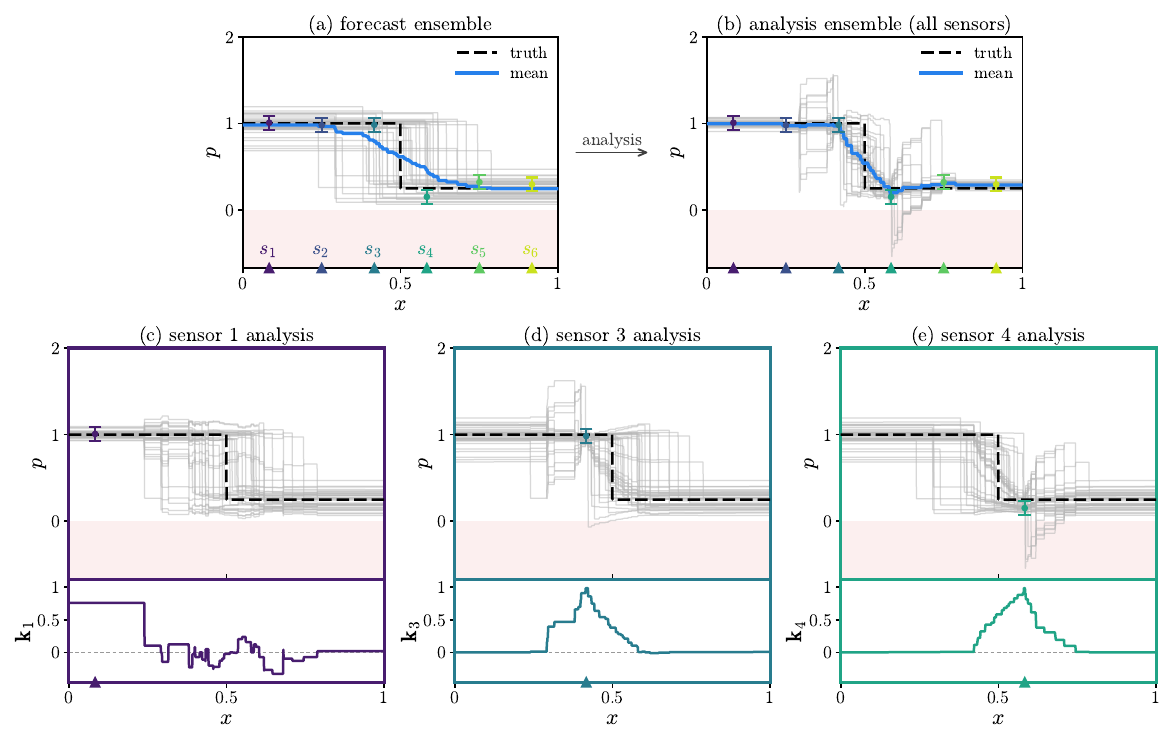}
\caption{Case 3 of Fig.~\ref{fig:physical-enkf-cases}, the misaligned-front ensemble: contributions of individual sensors to the EnKF update. (a) Forecast ensemble and observations. (b) Analysis ensemble using all six sensors. (c)-(e) Contributions of sensors $s_1$, $s_3$ and $s_4$ added separately to the forecast, each with the gain profile $\mathbf{k}_{\ell}$ of that sensor in the box beneath the panel.}
\label{fig:gain-profiles}
\end{figure}

Figure~\ref{fig:gain-scatter} compares the forecast regression relation and the corresponding analysis states for cases~1 and~3 using sensor~3 alone. In both cases, the forecast state is read at a generic location, $x_g=0.47$. The difference between cases~1 and~3 is the distribution of the forecast front locations. In case~1, the fronts are all aligned at x=0.5, so all ensemble members occupy the same pressure plateau at $x_g$. The forecast ensemble consequently forms a single cloud because sensor~3 and $x_g$ lie on the same left plateau for all members. Thus, a single fitted line provides an adequate description of the relation between the sensor measurement and the pressure at $x_g$. The analysis retains this single-group character, and the pointwise distribution remains unimodal. In case 3, however, the dispersed front locations cause different ensemble members to occupy different pressure plateaus at $x_g=0.47$. The forecast ensemble therefore separates into distinct groups in the sensor-state plane exhibiting bimodal probability marginals along both axes. The EnKF nevertheless represents the dependence of the pressure at $x_g$ on the sensor~3 measurement through a single linear regression coefficient computed from the complete ensemble. The issue is therefore the use of one linear relation for ensemble members belonging to different local populations. 

The slopes shown in Fig.~\ref{fig:gain-scatter} correspond only to Kalman gain at $x_g=0.47$ in relation to sensor $s_3$. Repeating the same regression at every grid point gives the spatial slope profile associated with each sensor, corresponding to a column of $\mathbf{K}$. Figures~\ref{fig:gain-profiles}(c)--(e) show the gain contributions from sensors 1, 3, and 4. Each profile extends well beyond the sensor location and across the region occupied by the dispersed forecast fronts, so the correction associated with a sensor acts over a finite spatial interval rather than only at the sensor location. The largest corrections arise from sensors located between the forecast and true front positions ($s_3$ and $s_4$), where the observation mismatch is largest. These corrections act over broad spatial intervals, and their combined effect lowers the pressure near the true front while the original forecast discontinuity remains near its displaced location. The resulting analysis therefore contains both the distributed correction and the surviving forecast jump, producing a non-monotone profile. The magnitude of each sensor contribution can be understood directly from the observation mismatch entering Eqn.~\eqref{eq:gain-columns}, as summarized in the following remark.

\begin{remark}
Consider a forecast member whose front is displaced from the true front. A sensor located between the two front positions samples different pressure plateaus in the forecast member and in the truth, so the corresponding observation mismatch, $\bigl[\mathbf{d}+\bm{\eta}_i-\mathbf{y}_i^{f}\bigr]_{\ell}$, is comparable to the plateau difference $\mu_L-\mu_R$. Multiplication by the gain profile $\mathbf{k}_{\ell}$ then distributes a correction of this magnitude over the spatial extent of the profile. By contrast, sensors for which the forecast member and the truth lie on the same plateau have much smaller observation mismatches and therefore contribute weaker corrections. This contrast in the sensor-wise corrections produces overshoots and undershoots in the updated state, leading to non-monotonicity.
\end{remark}

The presented gain-based interpretation is equivalent to the weighted-superposition view of Sec.~\ref{sec:superposition-view}. In Eqn.~\eqref{eq:affine}, the analysis member is expressed as an affine combination of forecast states with different front locations. Equation~\eqref{eq:gain-columns} describes the same update in terms of spatial corrections generated by the sensor mismatches. Both interpretations identify the same mechanism: uncertainty in front location generates correlations across the discontinuity, and the EnKF converts these correlations into distributed amplitude corrections over the region occupied by the forecast fronts. Because the discontinuity of each forecast member remains at its own location, these distributed corrections can combine with the displaced discontinuity to produce non-monotone analysis states. These findings motivate performing the EnKF update in a representation where sharp features, such as shocks, vary smoothly with the state coordinates.

\section{Feature-preserving latent-EnKF}
\label{sec:method}

\begin{figure}[!t]
\centering
\includegraphics[width=\linewidth]{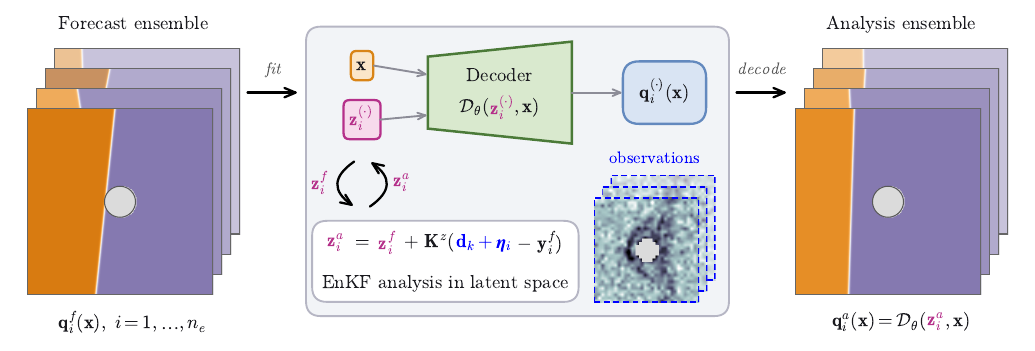}
\caption{One data assimilation cycle of the latent-EnKF. A shared CC-auto-decoder \cite{park2019deepsdf} represents each ensemble member as $\mathbf{q}_i^{(\cdot)}(\mathbf{x})=\mathcal{D}_{\bm{\theta}}(\mathbf{z}_i^{(\cdot)},\mathbf{x})$, with $(\cdot)=f$ or $a$. The forecast latent codes are updated using Eqn.~\eqref{eq:enkf-latent} and the noisy observations $\mathbf{d}_k+\boldsymbol{\eta}_i$, with latent-space gain $\mathbf{K}^{z}=\mathbf{Z}\mathbf{Y}^{\top}(\mathbf{Y}\mathbf{Y}^{\top}+\mathbf{R}_k)^{-1}$. Decoding the updated latent codes yields the analysis ensemble in physical space.}
\label{fig:latent-enkf}
\end{figure}

A feature-preserving latent-EnKF is proposed to circumvent the non-monotone updates generated by the physical-space EnKF. The proposed method performs the EnKF update in a low-dimensional latent representation learned by a CC-auto-decoder, rather than directly in physical space, with one assimilation cycle shown schematically in Fig.~\ref{fig:latent-enkf}. To construct this representation, each ensemble member of the forecast field is parameterized using a shared multilayer perceptron (MLP) decoder $\mathcal{D}_{\mathbf{\theta}}$ as
\begin{equation*}
\mathbf{q}_i^{f,k}(\mathbf{x})
\approx
\mathcal{D}_{\bm{\theta}}(\mathbf{z}_i^{f,k},\mathbf{x}).
\end{equation*}
Here, $\mathbf{z}_i^{f,k}\in\mathbb{R}^{n_z}$ denotes the latent code associated with ensemble member $\mathbf{q}_i^{f,k}$, with $n_z\ll n_q$. The decoder weights $\bm{\theta}$ and latent codes $\{\mathbf{z}_i^{f,k}\}_{i=1}^{n_e}$ are jointly optimized on the forecast ensemble $\{\mathbf{q}_i^{f,k}\}_{i=1}^{n_e}$ by minimizing
\begin{equation}
\label{eq:ad-loss-gw}
\mathcal{J}(\bm{\theta},\{\mathbf{z}_i\})
=
\underbrace{
\frac{1}{n_e}
\sum_{i=1}^{n_e}
\sum_{c=1}^{n_q}
\left\|
\omega_c(\mathbf{x})
\left[
\mathcal{D}_{\bm{\theta}}^{\,c}(\mathbf{z}_i,\mathbf{x})
-
q_{i,c}^{f}(\mathbf{x})
\right]
\right\|_{1}
}_{\text{gradient-weighted } L_{1} \text{ norm}}
+
\underbrace{
\frac{\beta}{n_z}
\sum_{j=1}^{n_z}
\left\|
\mathbf{z}_{(j)}
\right\|_{2}
}_{L_{2} \text{ Gaussian regularization}},
\end{equation}
with
\begin{equation*}
\omega_c(\mathbf{x})
=
\frac{1+\alpha\,\hat{g}_c(\mathbf{x})}
{\operatorname{mean}\left(1+\alpha\,\hat{g}_c\right)},
\qquad
\hat{g}_c
=
\frac{|\nabla q_c|}
{\max |\nabla q_c|}.
\end{equation*}
The symbols $\|\cdot\|_1$ and $\|\cdot\|_2$ denote the discrete $L_1$ and $L_2$ norms, respectively. The index $c$ denotes one of the $n_q$ flow variables, so that $q_c\in\{\rho,u,v,p\}$, and each component is reconstructed using its corresponding spatial weight $\omega_c$. The quantity $\hat{g}_c$ is computed from the gradient magnitude of component $c$ and normalized to $[0,1]$, with the maximum evaluated over the full forecast ensemble. The parameter $\alpha$ controls the additional reconstruction emphasis assigned to regions of large gradients; $\alpha=1$ is used throughout this work, while $\alpha=0$ reduces the reconstruction term to the standard $L_1$ norm. Normalization of $\omega_c$ by its spatial mean preserves its average magnitude while redistributing the reconstruction weight toward regions containing sharp flow features. The second term in Eqn.~\eqref{eq:ad-loss-gw} regularizes the latent ensemble. For latent coordinate $j$, $\mathbf{z}_{(j)}=(z_{1j},\ldots,z_{n_ej})$ collects that coordinate across all ensemble members. The regularization parameter is fixed at $\beta=10^{-4}$ in all experiments. The CC-auto-decoder provides a continuous and differentiable mapping from latent to physical space.

Before optimization, the spatial coordinates and flow variables are independently normalized to $[0,1]$. The decoder parameters $\bm{\theta}$ are initialized using the default Kaiming-uniform initialization in PyTorch, while the latent codes are initialized from $\mathcal{N}(\mathbf{0},\sigma_z^2\mathbf{I})$ with $\sigma_z=0.01$. The optimized forecast latent ensemble is denoted by $\{\mathbf{z}_i^{f}\}_{i=1}^{n_e}$. The EnKF analysis is then performed directly in the latent space. Define the latent forecast ensemble mean $\bar{\mathbf{z}}^{f}=n_e^{-1}\sum_i\mathbf{z}_i^{f}$, and the normalized perturbation matrix
\begin{equation}
\mathbf{Z}=\frac{1}{\sqrt{n_e-1}}\bigl[\mathbf{z}_i^{f}-\bar{\mathbf{z}}^{f}\bigr]_i\in\mathbb{R}^{n_z\times n_e}
\end{equation}
The latent codes are updated according to
\begin{equation}
\label{eq:enkf-latent}
\mathbf{z}_i^{a}=\mathbf{z}_i^{f}+\mathbf{Z}\mathbf{Y}^{\!\top}\!\bigl(\mathbf{Y}\mathbf{Y}^{\!\top}+\mathbf{R}_k\bigr)^{-1}\bigl(\mathbf{d}_k+\bm{\eta}_i-\mathbf{y}_i^{f}\bigr).
\end{equation}

The analysis state in physical space is recovered by decoding the updated latent code, $\mathbf{q}_i^{a,k}=\mathcal{D}_{\mathbf{\theta}}(\mathbf{z}_i^{a},\mathbf{x})$. Since all ensemble members share the same decoder parameters $\bm{\theta}$, variability in $\{\mathbf{z}_i^{f}\}$ directly represents variability in the underlying physical states. We emphasize that the EnKF analysis modifies only the latent codes in Eqn.~\eqref{eq:enkf-latent}; the decoder weights $\bm{\theta}$ remain fixed during the analysis step. At the next DA cycle, the decoder is re-trained on the updated forecast ensemble by warm-starting from the previously learned weights. Algorithm~\ref{alg:fs-enkf} summarizes series of steps in one DA cycle.

\begin{algorithm}[h!]
\caption{Feature-space ensemble Kalman filter (one DA cycle).}
\label{alg:fs-enkf}
\begin{algorithmic}[1]
\State Obtain forecast and predicted observations at $k^{\text{th}}$ DA cycle: $\mathbf{q}_i^{f} \gets \mathcal{F}_{\Delta t}(\mathbf{q}_i^{a,k-1})$, \quad $\mathbf{y}_i^{f} \gets \mathcal{H}(\mathbf{q}_i^{f})$
\State Fit the decoder $\mathcal{D}_{\mathbf{\theta}}$ and latent codes $\mathbf{z}_i^{f}$ to forecast state $\{\widetilde{\mathbf{q}}_i^{f}\}$ by minimizing Eqn.~\eqref{eq:ad-loss-gw}
\State Update forecast latent codes $\mathbf{z}_i^{f}$ via the EnKF analysis in Eqn.~\eqref{eq:enkf-latent} to obtain $\mathbf{z}_i^{a}$
\State Decode the posterior latent code to obtain the analysis state for the next forecast: $\mathbf{q}_i^{a,k} \gets \mathcal{D}_{\mathbf{\theta}}(\mathbf{z}_i^{a}, \mathbf{x})$.
\State \Return analysis ensemble $\{\mathbf{q}_i^{a,k}\}_{i=1}^{n_e}$
\end{algorithmic}
\end{algorithm}

\section{EnKF experiments}
\label{sec:experiments}

The proposed feature-preserving EnKF is demonstrated on two test cases in this section. The first is a Sod shock tube with sparse, noisy pressure observations. The second is a Mach 2 shock interacting with a two-dimensional cylinder with sparse, noisy observations proportional to $|\mathbf{\nabla}\rho|$. The second case, to our knowledge, is the first time that noisy density-gradient (schlieren-like) observations have been assimilated through the EnKF for shock-laden flows.

All experiments presented in this work employ a decoder architecture, $\mathcal{D}_{\boldsymbol{\theta}}$, consisting of five hidden layers of width 256 with ReLU activations. A skip connection is introduced by concatenating the inputs $[\mathbf{z}_i,\mathbf{x}_i]$ to the third hidden layer. Each ensemble member is represented by a 16-dimensional latent code $\mathbf{z}_i$, and no significant sensitivity is observed for latent dimensions larger than 16. At each DA cycle, a fresh decoder is trained on the forecast ensemble using Adam for 10,000 epochs with a learning rate of $10^{-3}$, mini-batches of 5,000 coordinates. No multiplicative or additive inflation is applied to the posterior latent ensemble.

\subsection{Sod shock tube}
\label{sec:sod}

\begin{figure}[h!]
\centering
\includegraphics[width=0.95\linewidth]{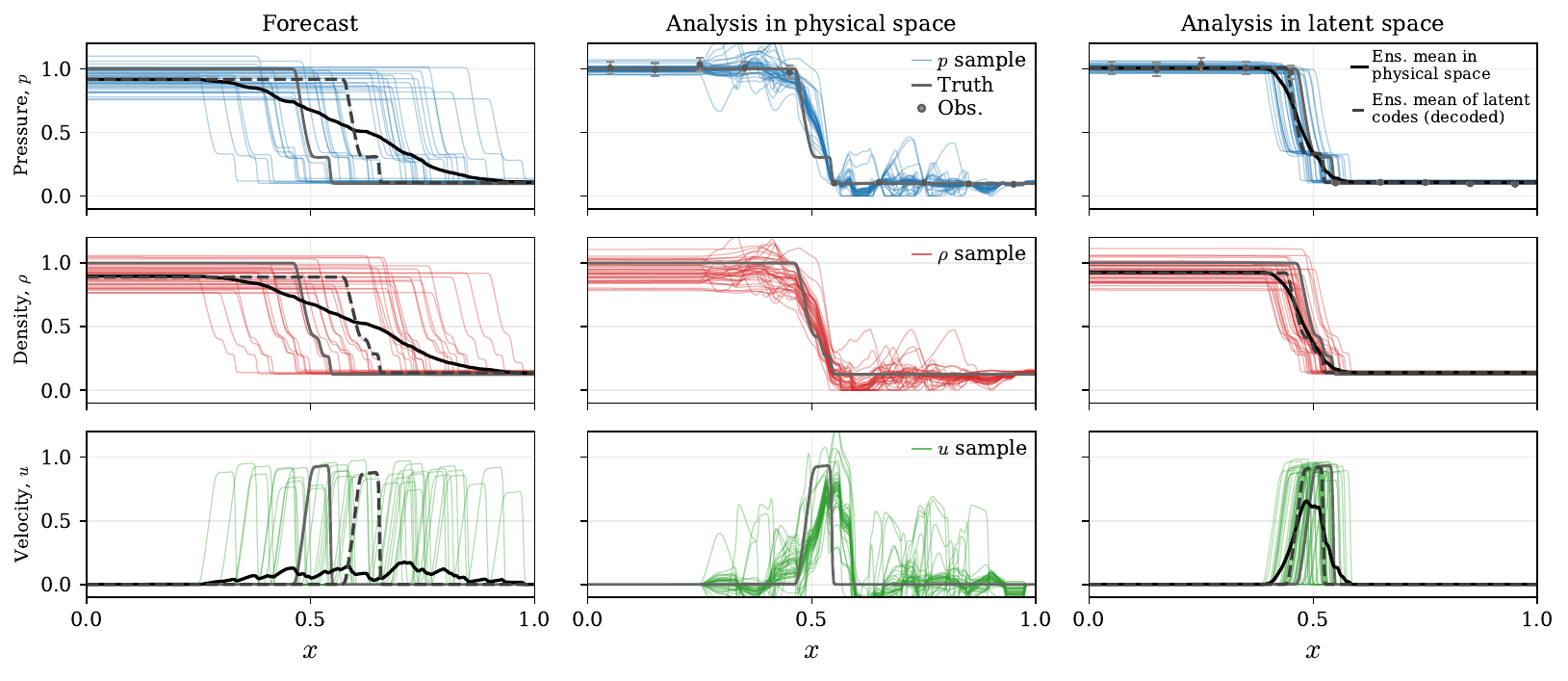}
\caption{Sod shock tube at the first DA cycle, $t=0.025$, with $p$, $\rho$, $u$ in the three rows. Left: forecast ensemble. Middle: analysis from the standard EnKF in physical space. Right: analysis from the feature-space latent-EnKF. The physical-space update produces oscillations near the shock, contact, and rarefaction, whereas the feature-space update contracts the ensemble onto the truth while preserving the sharp features.}
\label{fig:sod-t025}
\end{figure}

\begin{figure}[!t]
\centering
\includegraphics[width=0.95\linewidth]{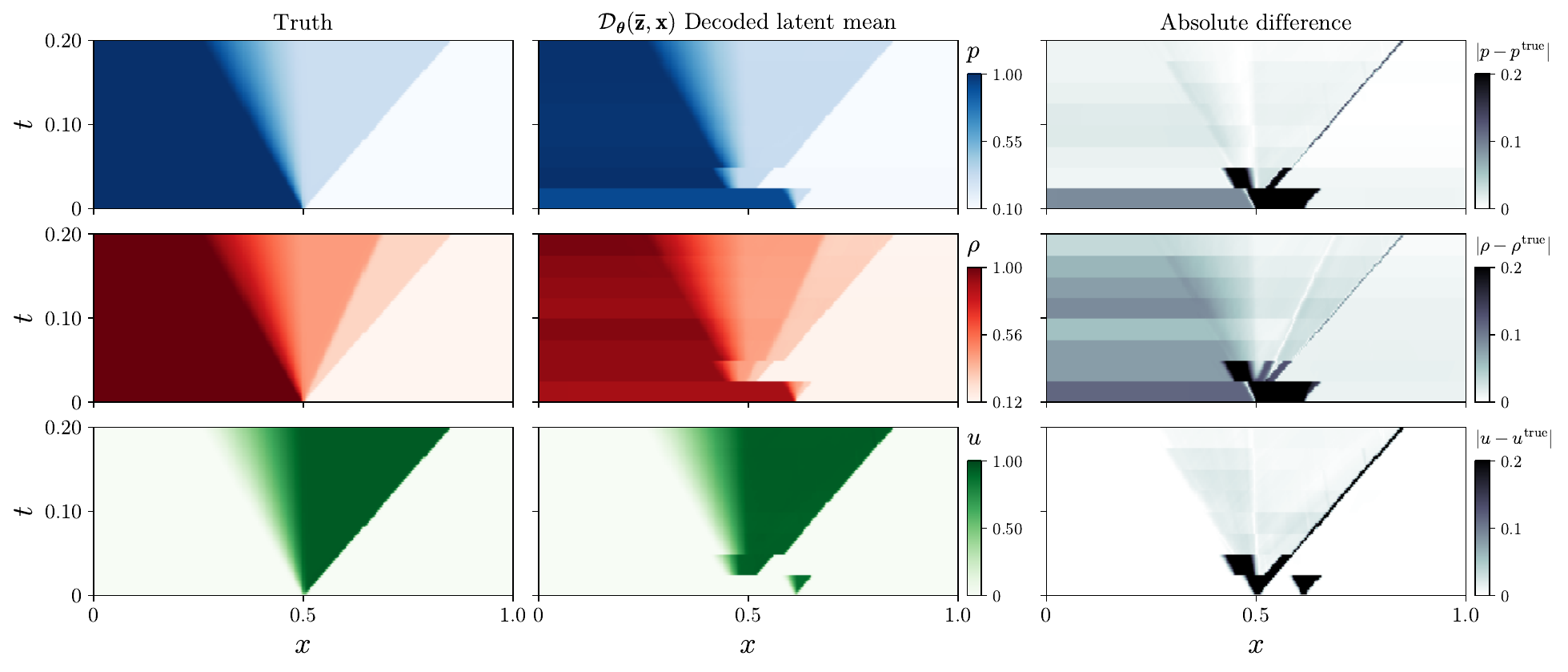}
\caption{ Space-time evolution of the Sod shock tube under latent-EnKF assimilation. Rows correspond to $p$, $\rho$, $u$. Left: analytical solution. Middle: forecast trajectory initialized from the decoded post-analysis latent mean $\mathcal{D}_{\bm{\theta}}(\bar{\mathbf{z}}^{a,k},\mathbf{x})$ at each DA cycle and advanced between observation times using the numerical solver. Right: absolute error relative to the truth}
\label{fig:sod-spacetime}
\end{figure}

\begin{figure}[t!]
\centering
\includegraphics[width=0.925\linewidth]{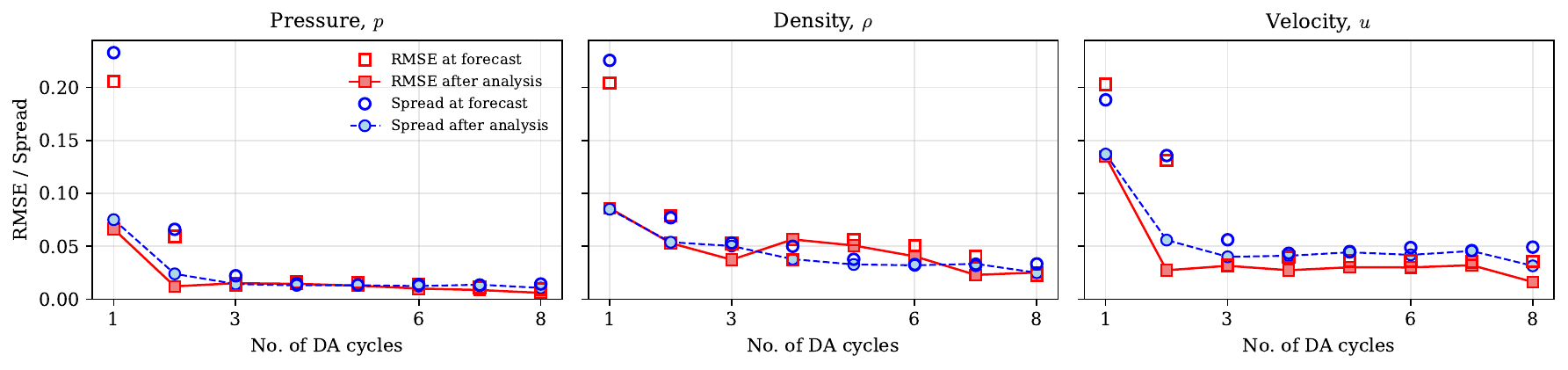}
\caption{Per-cycle RMSE (red squares) and ensemble spread (blue circles) for $p$, $\rho$, $u$ on the Sod test, evaluated at the forecast (open markers) and after the analysis (filled markers). Both metrics drop sharply at the first DA cycle and remain low thereafter.}
\label{fig:sod-rmse}
\end{figure}

The Sod shock tube problem is governed by the one-dimensional compressible Euler equations with specific-heat ratio $\gamma=1.4$. The canonical initial state $(\rho_L,u_L,p_L)=(1.0,0,1.0)$ and $(\rho_R,u_R,p_R)=(0.125,0,0.1)$ on $x\in[0,1]$ is separated by a diaphragm at $x_d=0.5$. Zero-gradient outflow boundary conditions are imposed at domain boundaries. The Euler equations are discretized on a uniform grid of $400$ cells using a WENO-5 reconstruction \cite{jiang1996efficient} with HLLC approximate Riemann solver for fluxes \cite{toro1994restoration, chandravamsi2023application}. The solution is advanced in time by a third-order SSP Runge-Kutta scheme \cite{gottlieb2001strong} to a final time $T=0.2$.

The initial-condition uncertainty is represented by an ensemble of $n_e=40$ members. The diaphragm location and the left/right primitive states are sampled independently from Gaussian distributions whose means are deliberately offset from the truth by $10\%$ for the left/right primitive states and $25\%$ for the diaphragm position, so that the prior is biased: $x_d\sim\mathcal{N}(0.625,0.20^2)$, $\rho_L\sim\mathcal{N}(0.9,0.10^2)$, $\rho_R\sim\mathcal{N}(0.1375,0.01^2)$, $p_L\sim\mathcal{N}(0.9,0.10^2)$, and $p_R\sim\mathcal{N}(0.11,0.01^2)$, with the corresponding truth values $0.5$, $1.0$, $0.125$, $1.0$, and $0.10$. Synthetic pressure observations are obtained by sampling the truth at $10$ uniformly spaced probes. The observation-noise covariance is heteroscedastic with an additive floor, $\mathbf{R}_k=\mathrm{diag}\bigl[(0.05\,|y_{k}^{\mathrm{true}}|+0.001)^{2}\bigr]$. The analysis is performed at eight DA cycles equi-spaced over $t\in[0.025,0.2]$ with $\Delta t_{\mathrm{obs}}=0.025$.

Fig.~\ref{fig:sod-t025} compares the forecast and analysis ensembles at the first DA cycle. The physical-space EnKF produces strong oscillations near the shock, contact discontinuity, and rarefaction in all primitive variables. Several ensemble members also develop negative density and pressure values after analysis, which are clipped to a positive floor of $10^{-3}$ in Fig.~\ref{fig:sod-t025} (middle column) to maintain thermodynamic admissibility. In contrast, the feature-space EnKF preserves sharp flow features while progressively aligning the ensemble toward the true solution. The space-time evolution in Fig.~\ref{fig:sod-spacetime} further shows that the forward trajectory initialized from the decoded post-analysis latent mean closely follows the analytical solution throughout the assimilation window, with residuals localized near propagating discontinuities. The slower density contraction observed in the per-cycle RMSE and ensemble spread (Fig.~\ref{fig:sod-rmse}) is consistent with the weak sensitivity of pressure observations to the contact discontinuity \cite{ke2026observability, zhou2026neural}.

\FloatBarrier

\subsection{Mach 2 shock-cylinder interaction}
\label{sec:cylinder}

\begin{figure}[t!]
\centering
\includegraphics[width=\linewidth]{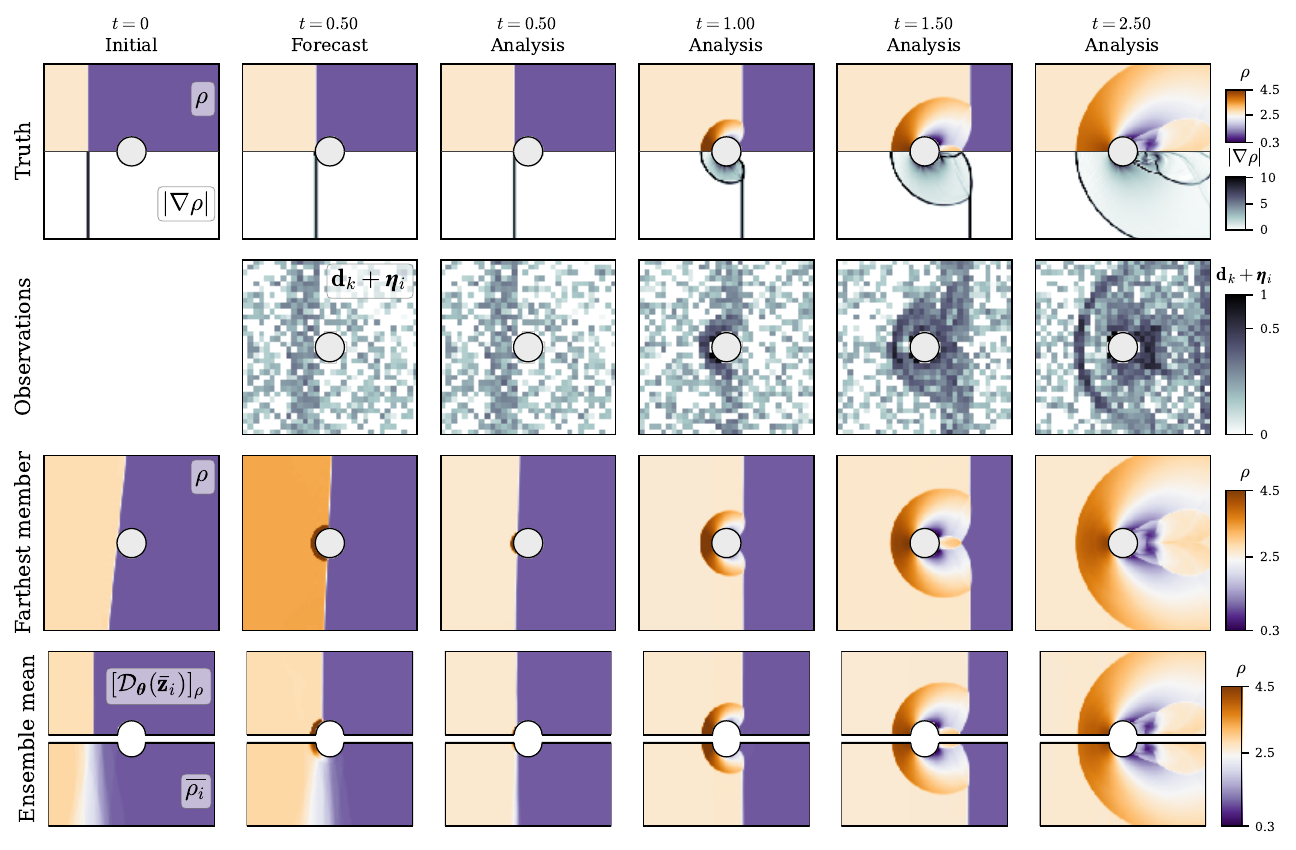}
\caption{Assimilation of Shock-cylinder interaction using latent-EnKF. Columns show the initial state, forecast, and analyses over successive DA cycles. Row 1: truth, with density $\rho$ (top half) and instantaneous $|\nabla\rho|$ (bottom half). Row 2: observations $\mathbf{d}_k +\boldsymbol{\eta}_i$, given by noisy, time-integrated $|\nabla\rho|$ on a $32\times32$ sensor grid. Row 3: farthest ensemble member from the mean at various stages of DA, illustrating ensemble spread. Row 4: Comparison between the ensemble mean obtained by decoding the mean latent code (top half) and the ensemble mean computed directly in the physical space (bottom half).}
\label{fig:cyl-grad}
\end{figure}

\begin{figure}[t!]
\centering
\includegraphics[width=0.9\linewidth]{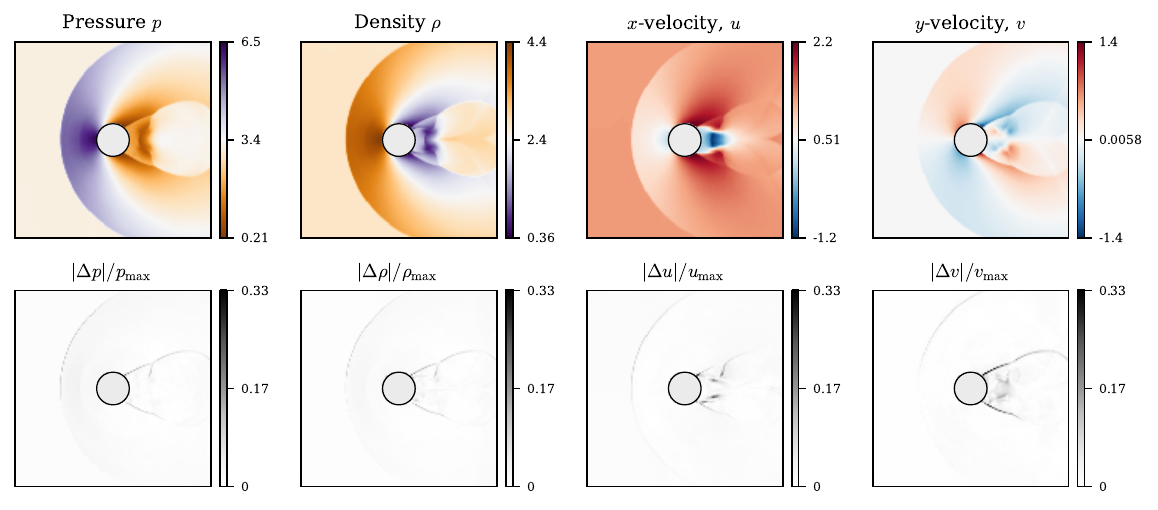}
\caption{Final-time ($t=2.5$) assimilated flow field for the cylinder case. Top row: decoded ensemble-mean primitive fields $p$, $\rho$, $u$, and $v$. Bottom row: absolute error relative to the truth, normalized by the maximum of each corresponding field, with a common color scale for direct comparison. Errors are localized near the bow shock, slip lines, and wake. The corresponding RMSE values are $\mathrm{RMSE}(p)\!\approx\!0.054$., $\mathrm{RMSE}(\rho)\!\approx\!0.037$, $\mathrm{RMSE}(u)\!\approx\!0.012$, and $\mathrm{RMSE}(v)\!\approx\!0.014$.}
\label{fig:cyl-final}
\end{figure}

\begin{figure}[t!]
\centering
\includegraphics[width=0.925\linewidth]{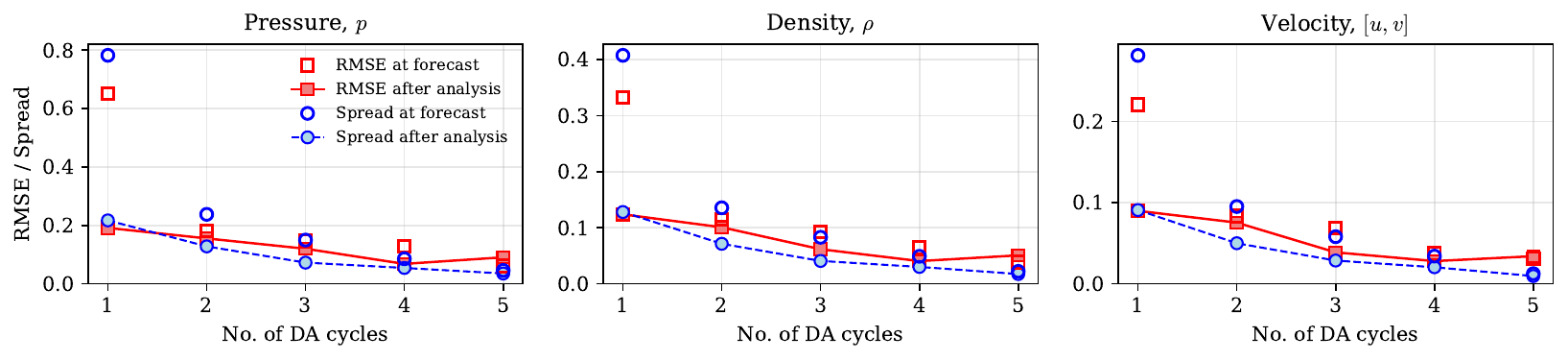}
\caption{RMSE and ensemble spread across DA cycles for the cylinder case with a $10\%$ biased prior. Results are shown for $p$, $\rho$, and velocity magnitude $|\boldsymbol{\upsilon}|$. Open markers denote forecast values, while filled markers denote post-analysis values. Both RMSE and spread decrease monotonically over five DA cycles, with the analysis spread closely following the RMSE.}
\label{fig:cyl-rmse}
\end{figure}

The second test case is the diffraction of a planar Mach-2 shock by a circular cylinder, governed by the two-dimensional compressible Euler equations with $\gamma=1.4$ on $\Omega=[0,L]^2$ with $L=6$. The pre-shock quiescent state is $(\rho,u,v,p)_1=(1,0,0,1/\gamma)$. The initial planar shock is located at $x_{\mathrm{s}}=L/4=1.5$ and propagates rightward at Mach $M=2$, with the post-shock state set by the Rankine-Hugoniot relations. A circular cylinder of diameter $D=1$ is centered at $(L/2,L/2)$ and represented by a sharp-interface immersed boundary with slip-wall ghost cells \cite{chaudhuri2011use}; zero-gradient (extrapolation) conditions are applied at all four domain boundaries. The spatial discretization and time integration are the same as in the Sod case, on a uniform $256\times 256$ grid at CFL $0.55$ to a final time $T=2.5$. The truth is obtained employing the aforementioned nominal initial conditions.

Uncertainty in the initial-condition is represented by an ensemble of $n_e=40$ members generated from Gaussian perturbations of three control parameters: the shock position $x_{\mathrm{s}}$, shock Mach number $M$, and shock tilt angle $\phi$. The inflow is rotated by the same $\phi$ to retain the flow normal to the shock. The ensemble members are generate with a biased mean, that is $10\%$ away from the truth in each control, with $x_{\mathrm{s}} \sim \mathcal{N}(1.65,\,0.40^2)$, $M \sim \mathcal{N}(2.20,\,0.20^2)$, and $ \phi \sim \mathcal{N}(0,\,0.20^2)$. The truth corresponds to $[x_{\mathrm{s}}, M, \phi]$ = $[1.5, 2.0, 0]$.

Observations are derived from the time-cumulative density-gradient magnitude over each DA interval, $\mathbf{d}_k(\mathbf{x})=\int_{t_{k-1}}^{t_k}|\nabla\rho|\,\mathrm{d}t$, which mimics a finite-time exposure in optical measurements. This integral is evaluated as a discrete sum over ten equally spaced accumulator sub-steps in time per DA cycle. The accumulated field is block-averaged onto a $32\times 32$ coarse sensor grid, obtained with a sparse factor of $8$ along each axis on the $N_x=N_y=256$ simulation grid. Sensors falling inside the cylinder are discarded, leaving $1000$ fluid sensors per cycle. The associated observation is denoted $\mathbf{d}_k+\boldsymbol{\eta}_i$, where $\boldsymbol{\eta}_i\sim\mathcal{N}(0,\mathbf{R}_k)$ and $\mathbf{R}_k$ is heteroscedastic with $\mathbf{R}_k=\mathrm{diag}\bigl[(0.12\,|y^{\text{true}}_{k}|+0.06)^2\bigr]$. Five DA cycles equispaced over $t\in[0,2.5]$ are used, with $\Delta t_{\mathrm{obs}}=0.5$.

Fig.~\ref{fig:cyl-grad} tracks the forecast ensemble at various DA cycles as the noisy, schlieren-like observations (row 2) extracted from truth (row 1) are assimilated. Row 3 follows the farthest forecast member, defined as the member with maximum $\|\mathbf{q}_i^{f}-\bar{\mathbf{q}}^{f}\|_2$ value, where $\bar{\mathbf{q}}^{f}=n_e^{-1}\sum_{i}\mathbf{q}_i^{f}$ is the primitive variable ensemble mean. The farthest member begins with a misplaced shock structure far from the truth, then progressively aligns with the truth as assimilation proceeds. Row 4 shows the decoded ensemble mean converging onto the truth, with the bow shock, Mach stem, and slip lines recovered cleanly. The full primitive state at $t=2.5$ (Fig.~\ref{fig:cyl-final}) is reconstructed in physical space without spurious oscillations across the bow shock. The EnKF update in latent space reconstructs a consistent shock and flow field (see Fig.~\ref{fig:sod-t025}). The largest residuals appear near the bow shock and the shock-cylinder interaction, consistent with the observation operator integrating $|\nabla\rho|$ over time, which provides limited information about transients, post shock-cylinder interaction. Fig.~\ref{fig:cyl-rmse} shows that RMSE and ensemble spread decrease monotonically over five DA cycles for all primitive variables, indicating that the latent-space update remains stable and informative.

\section{How the latent-EnKF resolves the failure of the physical-space EnKF}
\label{sec:mechanism}

Section~\ref{sec:experiments} showed that, unlike the physical-space EnKF, the latent-EnKF recovers shocks, contact discontinuities, and slip lines without generating spurious oscillations. This section examines the statistical properties of the latent-EnKF that enable this behavior. The discussion is organized into three parts, each revisiting the physical-space failure using one of the analyses developed in Sec.~\ref{sec:failure}. (i) Section~\ref{sec:mech-gauss} compares the bimodal pointwise statistics of the physical ensemble with the statistics of the corresponding latent codes; (ii) Section~\ref{sec:superpose} compares weighted superposition in physical and latent space; and (iii) Section~\ref{sec:mech-kernel} compares the regression interpretation of the Kalman gain in the two spaces.

\subsection{Gaussian statistics of the latent code}
\label{sec:mech-gauss}

The EnKF update of Eqn.~\eqref{eq:enkf-latent} depends on the mean and covariance of the forecast latent codes through the normalized ensemble perturbation matrix $\mathbf{Z}$. Section~\ref{sec:bimodal} showed that, in physical space, a point within the region occupied by the ensemble fronts samples either the pre-shock or post-shock state, giving a bimodal mixture. The latent code $\mathbf{z}_i$, in contrast, encodes the flow features of member $i$, while the decoder reconstructs the physical state from those features. The question is therefore whether the distributions of the latent coordinates remain close to Gaussian even when the corresponding pointwise physical-space statistics are bimodal.

\begin{figure}[!t]
\centering
\includegraphics[width=\linewidth]{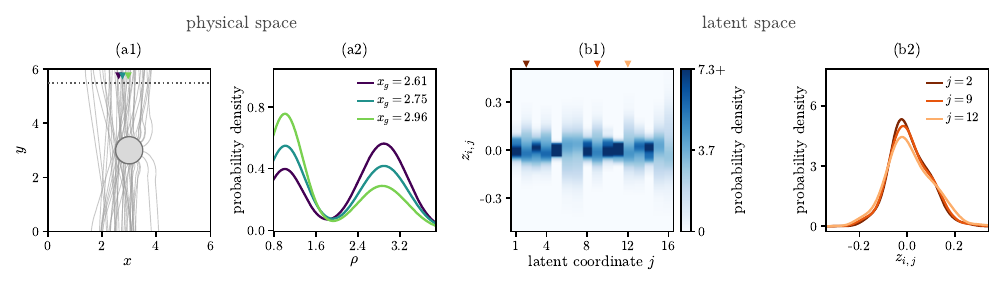}
\caption{Mach~2 shock-cylinder forecast ensemble at the first DA cycle, drawn separately with $n_e=128$ and $n_z=16$. (a1) Forty shock fronts, cylinder, and three points along $y=5.47$. (a2) Estimated probability density of $\rho$ at the three representative locations on $y=5.47$ shown in panel~(a1). (b1) Estimated probability density of each latent coordinate, with three selected. (b2) Estimated probability density of the selected coordinates. The distributions in panels~(a2), (b1), and~(b2) are obtained by fitting Gaussian kernels to each of the $n_e$ ensemble samples and averaging the resulting kernels, using a bandwidth of $n_e^{-1/5}$ times the sample standard deviation.}
\label{fig:latent-gauss-cylinder}
\end{figure}

Figure~\ref{fig:latent-gauss-cylinder} compares the pointwise physical-space statistics with the distributions of the latent coordinates for the shock-cylinder case of Sec.~\ref{sec:cylinder} at the first DA cycle, with $n_z=16$. The ensemble shown is a separate draw of $n_e=128$ members from the same prior, made larger so that the coordinate distributions are resolved. Panel~(a1) shows the ensemble shock fronts and three locations along $y=5.47$ indicated by upside down triangles on at $x_g=2.61$, $2.75$, and $2.96$, all within the region occupied by the fronts. For legibility, $40$ of the $128$ fronts are drawn, spanning the full spread of front positions. Panel~(a2) shows the estimated probability density of the mass density $\rho$ at these locations, with two peaks corresponding to the pre-shock and post-shock states. The relative amplitudes of the two peaks vary at these locations as the fraction of members on either side changes. Panel~(b1) shows the estimated probability density of each of the sixteen latent coordinates over the ensemble, with the coordinate index along the horizontal axis and the code value along the vertical axis. Three of these coordinates, $j=2$, $9$, and $12$, are selected, and panel~(b2) shows their estimated probability densities as curves.

The contrast between panels~(a2) and~(b2) follows from how the state values in the physical and latent representations vary with shock location. In physical space, the spread in shock location produces the bimodal pointwise statistics of panel~(a2), as discussed in Sec.~\ref{sec:bimodal}. The latent coordinates, however, vary smoothly with the ensemble controls, so nearby shock locations correspond to nearby latent codes rather than distinct groups. The role of latent coordinates in representing the different features of the ensemble is later discussed in Sec.~\ref{sec:representation}. The coordinate distributions in panel~(b2) therefore remain single-peaked, and none of the sixteen latent coordinates shows a statistically significant departure from a fitted normal distribution for the present ensemble size. The next subsection examines how the physical and latent representations affects weighted superposition of ensemble members.

\subsection{Physical vs.\ latent space superposition of ensemble members}
\label{sec:superpose}

In Section~\ref{sec:superposition-view}, we expressed each EnKF analysis member as a weighted superposition of the forecast members and showed that, in physical space, this superposition produces multiple undesirable discontinuities when the forecast fronts are misaligned. The same algebra applies to the latent-EnKF, but the weighted sum is formed in latent space before decoding. This difference is isolated here by comparing the same affine superposition of two states in physical and latent space.

\begin{figure}[t!]
\centering
\includegraphics[width=1.0\textwidth]{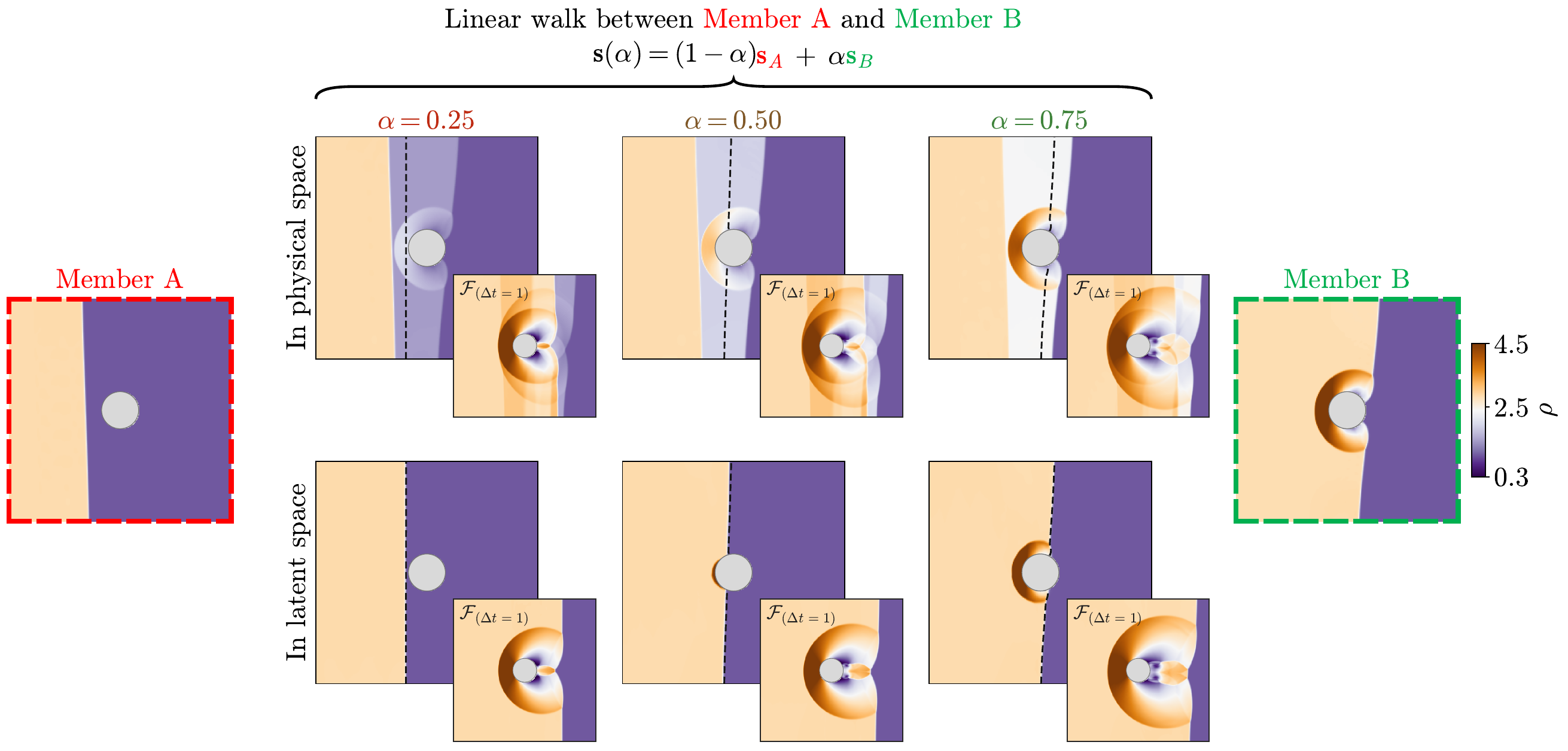}
\caption{Density $\rho$ of the Mach~2 shock-cylinder interaction at $t=0.5$. Members $A$ and $B$ are the forecast members whose shock fronts are farthest apart. Between them are the blends at $\alpha=0.25$, $0.50$ and $0.75$, formed in physical space (top row) and in the latent space of a CC-auto-decoder (bottom row). The dashed contour is the $\rho=1.5$ contour of the reference field. The inset in each  panel shows the corresponding state advanced by $\Delta t=1$ using the forecast model $\mathcal{F}_{\Delta t}$.}
\label{fig:superpose-cylinder}
\end{figure}

Equation~\eqref{eq:enkf-latent} has the same form as Eqn.~\eqref{eq:enkf-physical}, with $\mathbf{Z}$ replacing $\mathbf{Q}$ and the observation ensemble obtained from the decoded states. Therefore, the rearrangement of Sec.~\ref{sec:superposition-view} that led to Eqn.~\eqref{eq:affine},
\begin{equation}
\mathbf{q}_i^{a}
=
\sum_{j=1}^{n_e} w_{j}^{(i)}\mathbf{q}_j^{f},
\tag{\ref{eq:affine}}
\end{equation}
yields the corresponding latent-space form, 
\begin{equation}
\label{eq:affineLatent}
\mathbf{z}_i^{a}
=
\sum_{j=1}^{n_e} w_{j}^{(i)}\mathbf{z}_j^{f},
\end{equation}
with the same form of weights and unit row sum as in the physical-space analysis. The associated decoded state is
\begin{equation}
\mathbf{q}_i^{a}
=
\mathcal{D}_{\bm{\theta}}
\left(
\sum_j w_{j}^{(i)}\mathbf{z}_j^{f}.
\mathbf{x}
\right),
\end{equation}
The distinction between physical \eqref{eq:affine} and latent-space \eqref{eq:affineLatent} EnKF is therefore whether the weighted sum is taken before or after decoding. Figure~\ref{fig:superpose-cylinder} illustrates this difference using two members of the Mach~2 shock-cylinder forecast ensemble at the first DA cycle, $t=0.5$. Members $A$ and $B$ are the two whose shock fronts are farthest apart. For $\alpha=\{0.25, 0.50, 0.75\}$, the top row shows the physical-space superposition
\begin{equation}
\mathbf{q}(\alpha)
=
(1-\alpha)\mathbf{q}_A
+
\alpha\mathbf{q}_B,
\end{equation}
while the bottom row shows the decoded latent superposition
\begin{equation}
\mathbf{q}(\alpha)
=
\mathcal{D}_{\bm{\theta}}
\left(
(1-\alpha)\mathbf{z}_A
+
\alpha\mathbf{z}_B,
\mathbf{x}
\right).
\end{equation}
The dashed line is the correct shock location, obtained by simulating the flow form $\mathbf{c}(\alpha)=(1-\alpha)\mathbf{c}_{A}+\alpha\mathbf{c}_{B}$, where $\mathbf{c}_{A}$ and $\mathbf{c}_{B}$ are the control vectors from which members $A$ and $B$ are generated.

The training data of the CC-auto-decoder included four fields simulated from the control vectors at $\alpha=\{0, 1/3, 2/3, 1\}$, of which members $A$ and $B$ are the first and the last. The physical-space superposition retains one front from each member and produces an intermediate-density region between the two fronts. Around the cylinder, the physical-space superposition carries the diffraction pattern of member $B$ at reduced amplitude. The latent-space superposition instead produces a single front that moves continuously between members $A$ and $B$, with one diffraction pattern at the cylinder. The reference contour follows this motion across all three values of $\alpha$. The RMSE relative to the reference field is $\{0.383, 0.455, 0.405\}$ for the physical-space superpositions, compared with $\{0.033, 0.155, 0.046\}$ for the latent-space superpositions. Along the walk, $\alpha=0.50$ lies $1/6$ from the nearest fitted fields, whereas $\alpha=0.25$ and $0.75$ lie $1/12$ from their nearest fitted fields. The qualitative difference between physical- and latent-space superposition shown in Fig.~\ref{fig:superpose-cylinder} has direct implications for the subsequent EnKF forecast states. As seen in the insets of Fig.~\ref{fig:superpose-cylinder}, advancing the physical-space superpositions generates spurious wave transients that propagate through the domain, whereas the latent-space superpositions evolve without such artifacts.


The distinction between superposition in physical and latent space carries directly to the EnKF update. A physical-space affine superposition retains discontinuities at the locations present in the contributing forecast members, whereas a latent-space affine superposition is decoded as a state with a single front at an intermediate location. The same behavior is observed for the two-dimensional Riemann problem in \ref{app:riemann-walk}. The mechanism by which the decoder moves the front as the latent code varies is examined in Sec.~\ref{sec:mech-morph}.

\subsection{The Kalman kernel in latent and physical EnKF}
\label{sec:mech-kernel}

\begin{figure}[!t]
\centering
\includegraphics[width=1.0\textwidth]{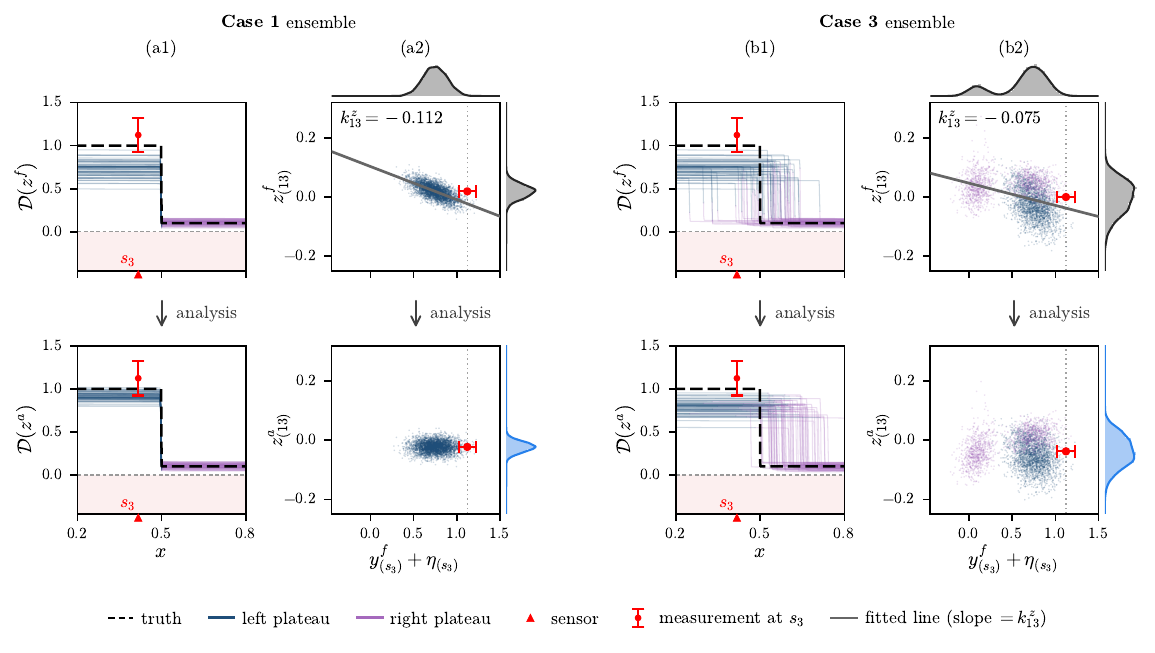}
\caption{Kalman-gain interpretation of the latent-EnKF update for cases~1 and~3. (a1,b1) Decoded forecast and analysis ensembles with sensor $s_3$ at $x_{(s_3)}=0.417$. (a2,b2) Latent coordinate $z_{(13)}$ plotted against the noisy measurement at $s_3$, with marginal probability densities. The fitted forecast slope gives the corresponding entry of the latent gain.}
\label{fig:gain-scatter-latent}
\end{figure}

The regression interpretation of the Kalman gain developed in Sec.~\ref{sec:gain} also applies to the latent-EnKF. The difference is that the state updated by the filter in latent-EnKF is the latent code rather than the physical field. For a single sensor, Eqn.~\eqref{eq:enkf-latent} becomes
\begin{equation}
\mathbf{z}_i^{a}
=
\mathbf{z}_i^{f}
+
\mathbf{k}^{z}\bigl(d+\eta_i-y_i^{f}\bigr),
\qquad
k^{z}_m
=
\frac{
\operatorname{Cov}\!\left(z_m^{f},\widetilde y^{f}\right)
}{
\operatorname{Var}\!\left(\widetilde y^{f}\right)
},
\end{equation}
where $k_m^{z}$ is the slope of the fitted line between latent coordinate $z_m$ and the noisy sensor measurement. Figure~\ref{fig:gain-scatter-latent} gives the latent-space counterpart of Fig.~\ref{fig:gain-scatter} for cases~1 and~3.

Figure~\ref{fig:gain-scatter-latent} uses the same case~1 and case~3 ensembles (of Sec.~\ref{sec:failure}), sensor location, observation, and observation perturbations as Fig.~\ref{fig:gain-scatter}, with the EnKF update performed in latent space. For case~1, both updates are described by a single ensemble cloud and a single fitted line, similar to the physical-space EnKF. For case~3, the distinction between the physical- and latent-space regression becomes evident. In physical space, the state at the generic point $x_g$ separates into the two plateau populations of Fig.~\ref{fig:gain-scatter}(d), so one fitted slope is applied across two distinct groups. The corresponding latent coordinate in Fig.~\ref{fig:gain-scatter-latent}(d) remains distributed as a single cloud, consistent with the latent-code statistics discussed in Sec.~\ref{sec:mech-gauss}. The latent gain therefore updates the code along the dependence represented by this cloud, and decoding the updated code retains a single front.

\begin{figure}[!t]
\centering
\includegraphics[width=0.95\textwidth]{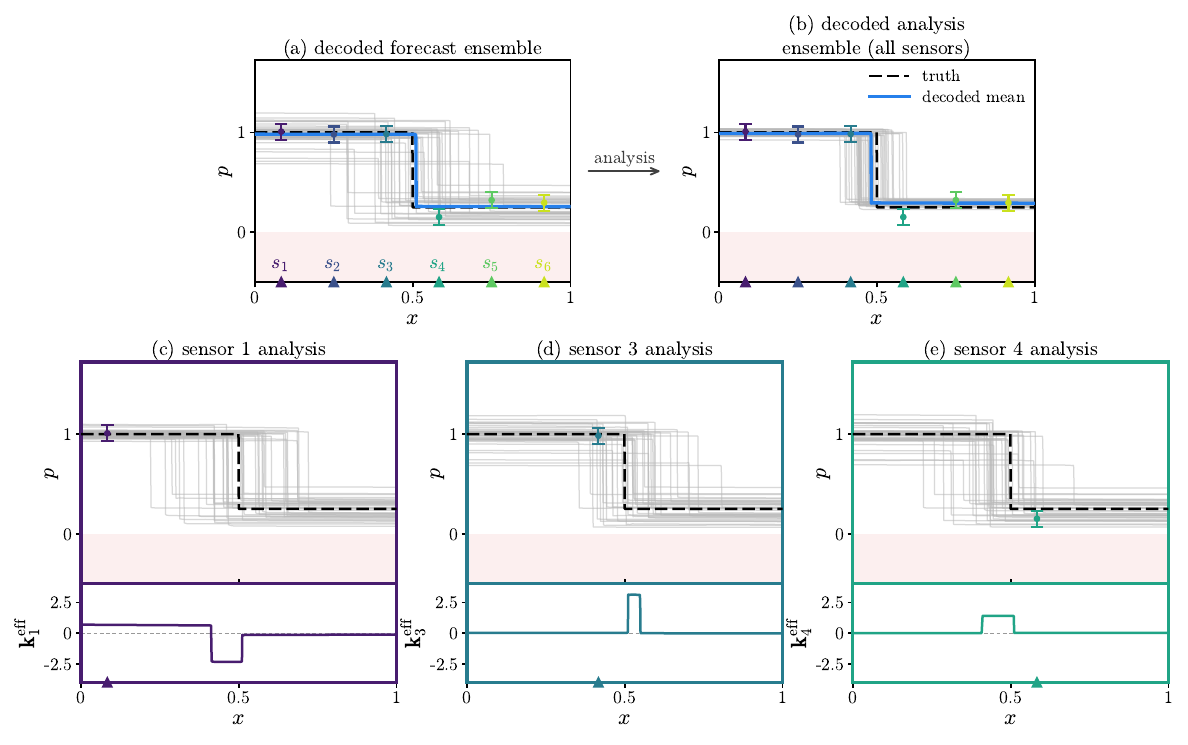}
\caption{Case 3 of Fig.~\ref{fig:physical-enkf-cases}, the misaligned-front ensemble: contributions of individual sensors to the latent-EnKF update. (a,b) Decoded forecast and analysis ensembles using all six sensors. (c)-(e) Decoded analysis obtained using sensors $s_1$, $s_3$ and $s_4$ separately, with the corresponding effective gain shown below.}
\label{fig:gain-profiles-latent}
\end{figure}

Figure~\ref{fig:gain-profiles-latent} shows how the latent gain directions appear after decoding back to physical space. Since $\mathbf{k}_{\ell}^{z}$ is defined in latent space, it has no direct interpretation in physical coordinates $\mathbf{x}$. The physical-space effect of $\mathbf{k}_{\ell}^{z}$ is represented through the decoder by
\begin{equation}
\label{eq:keff}
\begin{aligned}
t^{\mathrm{eff}}_{\ell}(x)
&=
\mathcal{D}_{\bm{\theta}}
\left(
\bar{\mathbf{z}}^{f}
+
\bigl[\mathbf{d}+\bm{\eta}_i-\mathbf{y}_i^{f}\bigr]_{\ell}
\mathbf{k}^{z}_{\ell},
x
\right)
-
\mathcal{D}_{\bm{\theta}}
\left(
\bar{\mathbf{z}}^{f},x
\right),\\
k^{\mathrm{eff}}_{\ell}(x)
&=
\frac{
t^{\mathrm{eff}}_{\ell}(x)
}{
\bigl[\mathbf{d}+\bm{\eta}_i-\mathbf{y}_i^{f}\bigr]_{\ell}
}.
\end{aligned}
\end{equation}

In Fig.~\ref{fig:gain-profiles-latent}(c-e), the effective latent kernels have an approximately square-wave character, with nearly constant corrections over the plateaus and a sharp transition near the front. This form is essential for preserving a discontinuous state during the update, because a smoothly varying kernel, such as a Gaussian-like profile, would apply different corrections to neighboring points and introduce artificial spatial variation into an otherwise piecewise-constant forecast state. The square-wave-like response instead modifies each plateau approximately uniformly while allowing the discontinuity to translate in space. The effective kernels nevertheless remain nonzero across much of the domain, and their relative plateau and front contributions depend on sensor location. Sensors that remain on the same plateau across the ensemble primarily adjust that plateau, whereas sensors within the region occupied by the forecast fronts act more strongly on the front location. This behavior reflects the two sources of variability in the forecast ensemble, namely uncertainty in the plateau states and in the front position.

\section{Representation mechanism of the coordinate-conditioned auto-decoder}
\label{sec:representation}

Section~\ref{sec:mechanism} showed that the latent-EnKF preserves near-Gaussian ensemble statistics, moves coherent flow features through latent-space updates, and yields physically consistent corrections after decoding. These properties arise from the representation learned by the CC-auto-decoder. This section examines how the CC-auto-decoder represents the flow field, the respective roles of the latent codes and shared decoder, and how this representation enters the EnKF update. The discussion is organized into three parts. Section~\ref{sec:mech-dof} examines how the latent degrees of freedom relate to the physical variations present across the forecast ensembles. Section~\ref{sec:mech-morph} examines coordinate morphing, namely whether variations in the latent code induce coordinate transformations that move shocks and geometric features through the decoder. Section~\ref{sec:mech-basis} then examines the basis functions represented by the last hidden layer and whether these basis functions remain shared across ensemble members as the flow features move.

\subsection{The latent code encodes the ensemble features}
\label{sec:mech-dof}

\begin{figure}[!t]
\centering
\includegraphics[width=0.95\textwidth]{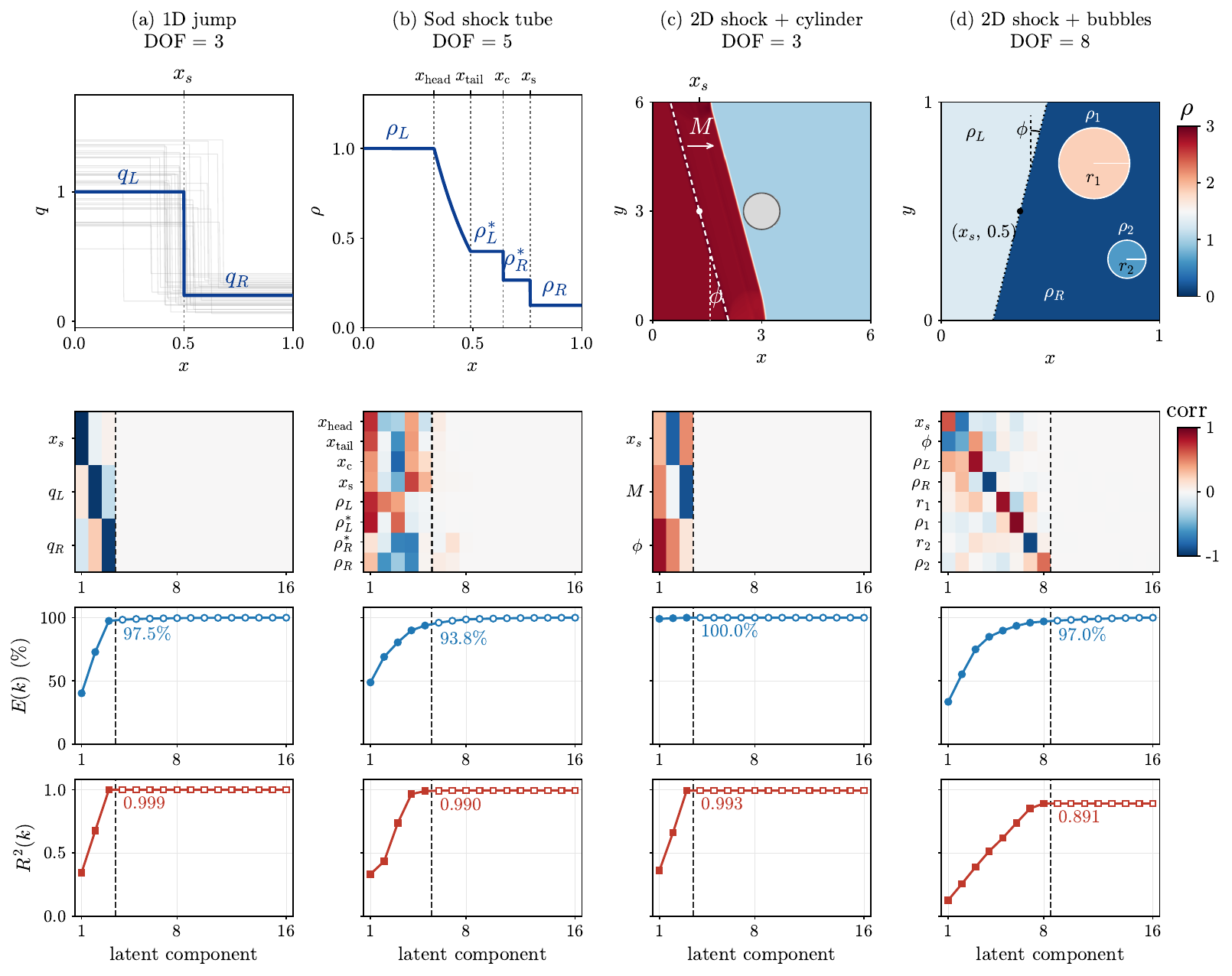}
\caption{Quantification of the correlation between the control parameters $\mathbf{c}$ prescribed for ensembles~(a)--(d) and the latent codes learned implicitly by the CC-auto-decoder. Row~1 shows a representative ensemble member together with the parameters varied across that ensemble. Row~2 shows the correlations between those parameters and the canonical latent directions defined by Eqn.~\eqref{eq:cca}. Row~3 shows the cumulative latent variance $E(k)$ from Eqn.~\eqref{eq:dof-energy}, and row~4 shows the cumulative $R^2(k)$ from Eqn.~\eqref{eq:dof-r2}, which measures how well the first $k$ canonical latent directions represent the prescribed control parameters.}
\label{fig:dof-cca-energy}
\end{figure}

The central question examined in this section is what physical information is represented by the latent code learned by the CC-auto-decoder. More specifically, we ask whether the latent code learned by the CC-auto-decoder encodes ensemble variations in physical features such as shock location, body position, body size, and other geometric or flow degrees of freedom. To aid the discussion, four simple ensembles (Fig.~\ref{fig:dof-cca-energy} row~1) are constructed with prescribed control parameters.
\begin{itemize}
    \item \textit{Ensemble (a):} A one-dimensional jump parameterized by the front position $x_s$ and plateau states $q_L$ and $q_R$, drawn independently from Gaussians truncated to the intervals shown, $x_s\sim\mathcal{N}(0.50,0.12^2)$ on $[0.12,0.88]$, $q_L\sim\mathcal{N}(1.00,0.20^2)$ on $[0.40,1.60]$, and $q_R\sim\mathcal{N}(0.20,0.10^2)$ on $[0,0.50]$.

    \item \textit{Ensemble (b):} The Sod shock tube of Sec.~\ref{sec:sod}, parameterized by the diaphragm position $x_d$ and the states $\{\rho_L, \rho_R, p_L, p_R\}$, with $x_d\sim\mathcal{U}(0.40,0.55)$, $\rho_L\sim\mathcal{U}(0.80,1.20)$, $\rho_R\sim\mathcal{U}(0.10,0.18)$, $p_L\sim\mathcal{U}(0.80,1.30)$, and $p_R\sim\mathcal{U}(0.08,0.15)$. The draws are repeated until the wave pattern lies inside the domain, at the time of the snapshot. The additional locations marked in the panel are derived features of the resulting wave pattern and are therefore functions of these five parameters.

    \item \textit{Ensemble (c):} The Mach~2 shock-cylinder problem of Sec.~\ref{sec:cylinder}, parameterized by shock position $x_s$, Mach number $M$, and inclination $\theta$, with $x_s\sim\mathcal{N}(1.65,0.40^2)$, $M\sim\mathcal{N}(2.20,0.20^2)$, and $\theta\sim\mathcal{N}(0,0.20^2)$, the prior of Sec.~\ref{sec:cylinder}.

    \item \textit{Ensemble (d):} A tilted two-dimensional shock with two bubbles, parameterized by $x_s$, $\theta$, $\rho_L$, $\rho_R$, and the radius and density $(r_1,\rho_1)$ and $(r_2,\rho_2)$ of the two bubbles, with $x_s\sim\mathcal{U}(0.35,0.65)$, $\theta\sim\mathcal{U}(-0.25,0.25)$, $\rho_L\sim\mathcal{U}(0.8,1.4)$, $\rho_R\sim\mathcal{U}(0.1,0.5)$, $r_1\sim\mathcal{U}(0.12,0.17)$, $\rho_1\sim\mathcal{U}(1.6,2.6)$, $r_2\sim\mathcal{U}(0.08,0.11)$, and $\rho_2\sim\mathcal{U}(0.50,0.70)$.
\end{itemize}

Each ensemble is learned with a separate CC-auto-decoder using a latent dimension of $n_z=16$. The architecture and hyperparameters are the same as those used in Sec.~\ref{sec:experiments}, except that training is carried out for four times as many epochs. To compare the prescribed physical parameters of the ensemble and the variations in the latent codes, we evaluate
\begin{equation}
\label{eq:cca-data}
\mathbf{C}'=\bigl[\mathbf{c}_j-\bar{c}_j\bigr]_j^{\!\top}\in\mathbb{R}^{n_c\times n_e},
\qquad
\mathbf{Z}'=\bigl[\mathbf{z}_i^{f}-\bar{\mathbf{z}}^{f}\bigr]_i\in\mathbb{R}^{n_z\times n_e},
\end{equation}
where each row of $\mathbf{C}'$ contains one centered control parameter across the ensemble, and each column of $\mathbf{Z}'$ contains the centered latent code of one ensemble member. The thin singular value decompositions of $\mathbf{C}'$ and $\mathbf{Z}'$ are
\begin{equation}
\label{eq:cca-svd}
\mathbf{C}'=\mathbf{U}_C\bm{\Sigma}_C\mathbf{V}_C^{\!\top},
\qquad
\mathbf{Z}'=\mathbf{U}_Z\bm{\Sigma}_Z\mathbf{V}_Z^{\!\top},
\qquad
\bm{\Sigma}_Z=\operatorname{diag}(\sigma_1,\dots,\sigma_{n_z}),
\end{equation}
so that $\mathbf{V}_C$ and $\mathbf{V}_Z$ provide orthonormal bases for the control and latent subspaces across the ensemble members.

Given the orthogonal bases $\mathbf{V}_C$ and $\mathbf{V}_Z$ of the control and latent spaces, canonical correlation analysis (CCA) \cite{hotelling1936relations,bjorck1973numerical} identifies the latent directions that align most closely with the control subspace. The singular value decomposition
\begin{equation}
\label{eq:cca}
\mathbf{V}_C^{\!\top}\mathbf{V}_Z
=
\widetilde{\mathbf{U}}\,
\widehat{\bm{\Lambda}}\,
\widetilde{\mathbf{V}}^{\!\top},
\qquad
\widehat{\bm{\Lambda}}\in\mathbb{R}^{n_c\times n_z},
\end{equation}
gives the canonical correlations $1\ge\hat{\lambda}_1\ge\cdots\ge\hat{\lambda}_d\ge0$, where $d=\min(n_c,n_z)$. Geometrically, each $\hat{\lambda}_j$ measures the alignment between a canonical direction in the control subspace spanned by $\mathbf{V}_C$ and the corresponding canonical direction in the latent subspace spanned by $\mathbf{V}_Z$. A value of $\hat{\lambda}_j=1$ corresponds to a perfect alignment between the corresponding control and latent directions, while smaller values correspond to weaker alignment. Rotating the latent basis $\mathbf{V}_Z$ by the orthogonal matrix $\widetilde{\mathbf{V}}$ gives
\begin{equation}
\label{eq:cca-variates}
\mathbf{B}
=
\mathbf{V}_Z\widetilde{\mathbf{V}}
=
\bigl[\mathbf{b}_1,\dots,\mathbf{b}_{n_z}\bigr]
\in\mathbb{R}^{n_e\times n_z},
\end{equation}
where the columns $\mathbf{b}_k$ are the canonical latent directions. The first $d$ columns of $\mathbf{B}$ are ordered by decreasing canonical correlation $\{\hat{\lambda}_k\}_{k=1}^{d}$. Thus, $\mathbf{b}_1$ has the strongest alignment with the control subspace, with subsequent directions having progressively weaker alignment. The correlation between control $j$ and canonical latent direction $k$ is
\begin{equation}
\label{eq:dof-corr}
M_{jk}
=
\frac{
\mathbf{c}_j^{\!\top}\mathbf{b}_k
}{
\sqrt{\mathbf{c}_j^{\!\top}\mathbf{c}_j}\,
\sqrt{\mathbf{b}_k^{\!\top}\mathbf{b}_k}
},
\qquad
j=1,\dots,n_c,
\quad
k=1,\dots,n_z,
\end{equation}
and is shown in row~2 of Fig.~\ref{fig:dof-cca-energy}.

The effective dimensionality of the latent representation is quantified using two complementary measures. (i) The fraction of latent variance carried by the first $k$ singular directions of $\mathbf{Z}'$,
\begin{equation}
\label{eq:dof-energy}
E(k)
=
\frac{
\sum_{i=1}^{k}\sigma_i^{2}
}{
\sum_{i=1}^{n_z}\sigma_i^{2}
},
\end{equation}
shown in row~3. (ii) The extent to which the first $k$ canonical latent directions represent the prescribed physical parameters, quantified by
\begin{equation}
\label{eq:dof-r2}
R^{2}(k)
=
\frac{1}{n_c}
\sum_{j=1}^{n_c}
\left(
1-
\frac{
\bigl\|\mathbf{c}'_j-\widehat{\mathbf{c}}_j^{(k)}\bigr\|_{2}^{2}
}{
\bigl\|\mathbf{c}'_j\bigr\|_{2}^{2}
}
\right),
\end{equation}
shown in row~4, where
\begin{equation}
\label{eq:dof-fit}
\widehat{\mathbf{c}}_j^{(k)}
=
\sum_{l=1}^{k}
\bigl(\mathbf{b}_l^{\!\top}\mathbf{c}'_j\bigr)\,
\mathbf{b}_l
\end{equation}
is the orthogonal projection of the centered control $\mathbf{c}'_j=\mathbf{c}_j-\bar{c}_j$ onto the first $k$ canonical latent directions.

Row~2 of Fig.~\ref{fig:dof-cca-energy} shows that the prescribed physical parameters are correlated mainly with the first $n_c$ canonical latent directions, while the remaining directions have negligible correlation. Row~3 shows a similar result from the latent variance. For ensembles~(a) to (d), respectively, the first $\{3, 5, 3, 8\}$ singular directions contain $\{97.8, 94.2, 90.1, 97.6\}\%$ of the latent variance . In each ensemble, most of the latent variation is therefore carried by the same number of directions as the number of independently varied physical parameters.

For ensembles~(a) to (d), respectively, row~4 shows that the cumulative $R^2$ reaches $\{0.999, 0.989, 0.999, 0.991\}$ at $k=\{3, 5, 3, 8\}$. In every case, the number of canonical latent directions required to recover the prescribed physical parameters matches the number of independently varied parameters in the ensemble. The results from row~4 therefore show that the active latent subspace closely matches the subspace spanned by the prescribed physical parameters. Thus, an EnKF update performed in the latent coordinates closely corresponds to performing the EnKF update directly on the physical controls. In the assimilation problems of Sec.~\ref{sec:experiments}, these controls are not supplied explicitly; they are represented implicitly by the latent code learned from the forecast ensemble. Section~\ref{sec:mech-morph} examines how variations in these latent coordinates are mapped by the decoder into changes in the physical flow features.

\subsection{Latent coordinate morphing: the latent code transforms the coordinates read by the shared decoder}
\label{sec:mech-morph}

Section~\ref{sec:superpose} showed that weighted superposition of members in latent space preserves the geometric form of the flow features while continuously morphing one flow realization into another. Such behavior suggests that the latent code does not merely alter the reconstructed state directly, but instead modifies the coordinates through which the shared decoder represents the flow. Under this interpretation, variations in shock position, body position, and body size are encoded through a latent-dependent coordinate transformation, while the downstream decoder remains common to all ensemble members and reconstructs the flow field from the transformed coordinates. The coordinate-morphing hypothesis can be written as
\begin{equation}
\mathcal{D}_{\bm{\theta}}(\mathbf{z},\mathbf{x})
=
\mathcal{G}_{\bm{\tilde{\theta}}}
\left(
\mathcal{T}(\mathbf{z},\mathbf{x})
\right),
\qquad
\tilde{\mathbf{x}}
=
\mathcal{T}(\mathbf{z},\mathbf{x}),
\end{equation}
where $\mathcal{T}$ denotes a latent-dependent mapping from the physical coordinates $\mathbf{x}$ to the internal coordinates $\tilde{\mathbf{x}}$, and $\mathcal{G}_{\bm{\tilde{\theta}}}$ denotes the shared downstream decoder. If this interpretation is correct, corresponding flow features from different ensemble members should approach common locations in $\tilde{\mathbf{x}}$ even when they occupy different locations in physical space.

\begin{figure}[!t]
\centering
\includegraphics[width=0.95\textwidth]{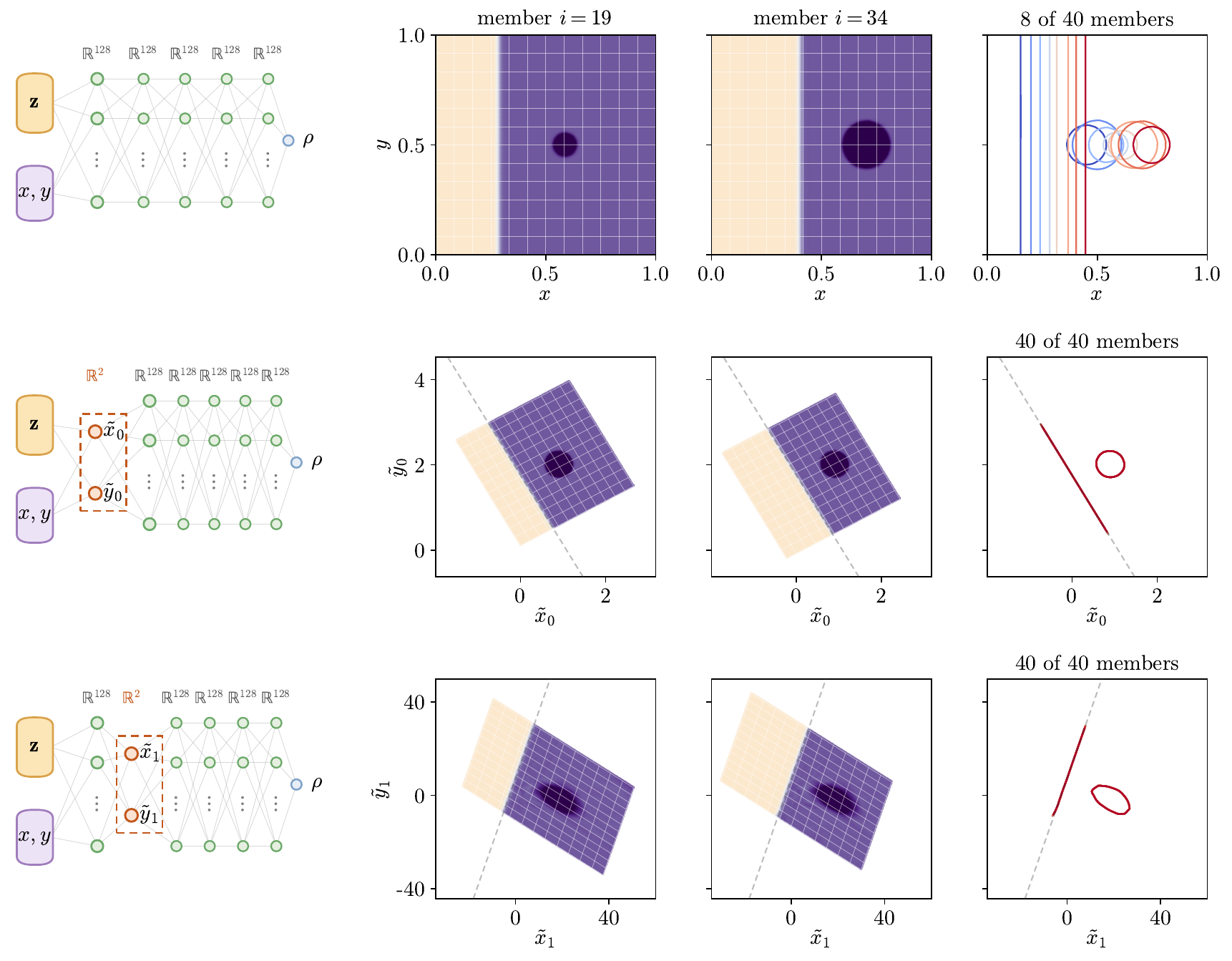}
\caption{Shock-cylinder ensemble of Sec.~\ref{sec:cylgeom} viewed through a two-neuron layer inserted into the frozen trained decoder. Column~1 shows the decoder architecture, columns~2 and~3 show members $i=19$ and $34$, and column~4 overlays the shock and cylinder contours from all ensemble members. Rows~1 to~3 show the fields in $(x,y)$, $(\tilde{x}_{0},\tilde{y}_{0})$, and $(\tilde{x}_{1},\tilde{y}_{1})$, respectively. Density scale as in Fig.~\ref{fig:superpose-cylinder}.}
\label{fig:morphing-2d}
\end{figure}

In order to test the above idea, a two-neuron linear layer is inserted into the trained decoder to examine whether the member-dependent variation can be represented through a two-dimensional internal coordinate system. The CC-auto-decoder used for this test has five hidden layers of width $128$ and latent dimension $n_z=16$, and is trained on $40$ members of the shock-cylinder configuration of Sec.~\ref{sec:cylgeom}. The inputs $[\mathbf{z}_i,\mathbf{x}_i]$ enter this decoder at the first layer alone. The cylinder position and diameter vary across the ensemble, while the shock remains a fixed distance $d=0.30$ upstream of the cylinder center.

The two-neuron layer is inserted at two depths, indexed by $m=0$ and $1$, corresponding to locations before and after the first hidden layer, respectively. Let $\mathbf{h}_{m}$ denote the representation entering the inserted layer, with
\begin{equation}
\mathbf{h}_{0}
=
\begin{bmatrix}
\mathbf{z}\\
\mathbf{x}
\end{bmatrix},
\qquad
\mathbf{h}_{1}
=
\mathrm{ReLU}
\left(
\mathbf{W}_{1}
\begin{bmatrix}
\mathbf{z}\\
\mathbf{x}
\end{bmatrix}
+
\mathbf{b}_{1}
\right).
\end{equation}
For either insertion, the transformed coordinates are defined as
\begin{equation}
\tilde{\mathbf{x}}_{m}
=
\begin{bmatrix}
\tilde{x}_{m}\\
\tilde{y}_{m}
\end{bmatrix}
=
\mathbf{B}_{m}\mathbf{h}_{m},
\qquad m\in\{0,1\},
\label{eq:morph-map}
\end{equation}
and the decoder can be written as
\begin{equation}
\mathcal{D}_{\bm{\theta}}(\mathbf{z},\mathbf{x})
=
\mathcal{G}_{\bm{\tilde{\theta}},m+1}
\left(
\mathbf{A}_{m}\tilde{\mathbf{x}}_{m}
\right),
\qquad m\in\{0,1\},
\label{eq:morph-split}
\end{equation}
where $\mathcal{G}_{\bm{\tilde{\theta}},m+1}$ denotes the frozen portion of the trained decoder downstream of the inserted layer. Thus, for each insertion, all dependence on the ensemble member and spatial location is restricted to pass through the two-dimensional variable $\tilde{\mathbf{x}}_{m}$ before reaching the remaining decoder layers.

All original decoder weights, biases, and latent codes are held fixed. Only the matrices $\mathbf{B}_{m}$ and $\mathbf{A}_{m}$ associated with the inserted two-neuron bottleneck layer are optimized. Their initial values are obtained from the rank-two truncated singular value decomposition of the frozen weight matrix replaced by the insertion, providing the optimal rank-two approximation in the least-squares sense. The matrices are subsequently optimized using the same training data, spatial grid, loss weighting, and optimization parameters as the original decoder. The procedure therefore tests whether the member-dependent variation already represented by the trained network admits a two-dimensional internal coordinate representation, without altering the original decoder parameters or latent codes.

Figure~\ref{fig:morphing-2d} compares the shock-cylinder ensemble in physical and transformed coordinates. In physical space, the shock location, cylinder position, and cylinder diameter vary across the ensemble. In $(\tilde{x}_{0},\tilde{y}_{0})$, the shocks collapse onto a common line and the cylinder boundaries onto a common curve, and this alignment is retained in $(\tilde{x}_{1},\tilde{y}_{1})$. The collapse of different ensemble members shows that the latent code maps different physical realizations to common internal coordinates before the downstream decoder reconstructs the flow. Because the alignment already appears for $m=0$, before the first nonlinear activation, the ensemble variation is largely represented through the two latent-dependent coordinates. The latent code therefore controls the internal coordinates at which the shared downstream decoder evaluates the flow features, explaining the continuous feature motion observed under latent interpolation in Sec.~\ref{sec:superpose}. Section~\ref{sec:mech-basis} examines the corresponding feature representation at the final hidden layer of the decoder.

\subsection{Basis functions of the shared decoder}
\label{sec:mech-basis}

\begin{figure}[!t]
\centering
\includegraphics[width=0.85\linewidth]{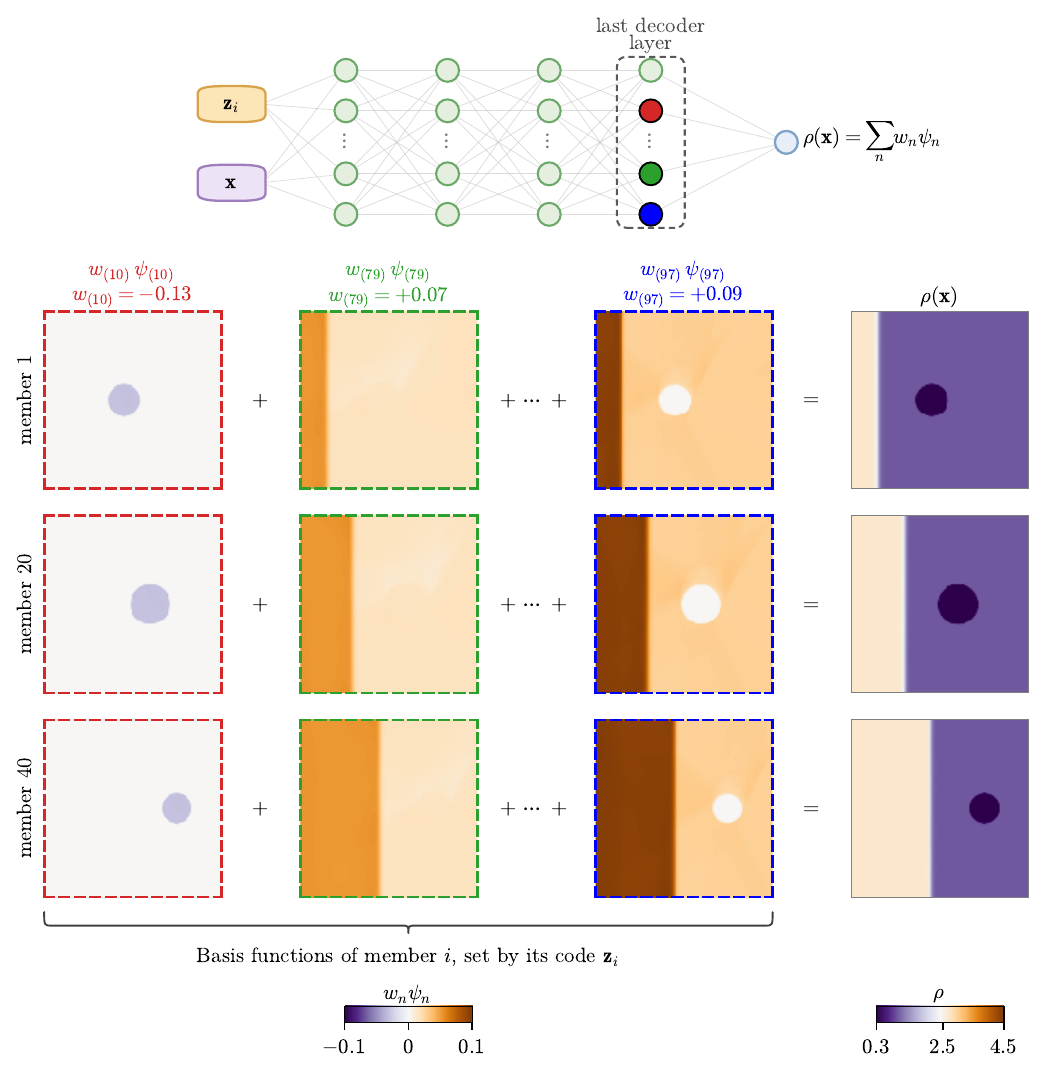}
\caption{Decoded density field of the shock-cylinder ensemble of Sec.~\ref{sec:cylgeom} expressed as the weighted sum of Eqn.~\eqref{eq:basis-sum}. The top row shows the decoder and its final hidden layer. The remaining rows correspond to members $i=1$, $20$, and $40$; the three panels left of the equals sign show the contributions $w_n\psi_n$ from units $\psi_{(10)}$, $\psi_{(79)}$, and $\psi_{(97)}$, while the ellipsis denotes the remaining terms and the rightmost panel shows the decoded field.}
\label{fig:basis-2d}
\end{figure}

The coordinate-morphing interpretation of Sec.~\ref{sec:mech-morph} can also be examined from the output side of the decoder. The decoded density field can be expressed as a weighted sum of the post-activation states in the last hidden layer,
\begin{equation}
\label{eq:basis-sum}
\rho(\mathbf{x})
=
\sum_n
w_n\,\psi_n(\mathbf{z}_i,\mathbf{x}),
\end{equation}
where $\psi_n$ is the post-activation state of the $n^{\mathrm{th}}$ unit in the final hidden layer, evaluated for member $i$ at coordinate $\mathbf{x}$, and $w_n$ is its output weight. Because the same decoder is used for every member, the weights $w_n$ are shared across the ensemble. The member dependence enters through $\psi_n(\mathbf{z}_i,\mathbf{x})$. Figure~\ref{fig:basis-2d} illustrates this decomposition for three members of the shock-cylinder ensemble of Sec.~\ref{sec:mech-morph}. Three representative terms are shown: $w_{(10)}\psi_{(10)}$, $w_{(79)}\psi_{(79)}$, and $w_{(97)}\psi_{(97)}$, together with the final decoded density field.

The individual terms reveal that the last hidden layer carries recognizable flow features. The contribution from $\psi_{(10)}$ is localized around the body, $\psi_{(79)}$ contains the shock front, and $\psi_{(97)}$ contains contributions associated with both features. Across the three members, these features follow the corresponding body and shock locations while retaining nearly the same spatial form. The decoder therefore uses a common set of output features across the ensemble, while the latent code determines their member-specific locations and amplitudes. This behavior is consistent with the coordinate transformation observed in Sec.~\ref{sec:mech-morph}: changes in the code alter where the shared decoder features appear in physical space.

The shared-feature interpretation demonstrated in Fig.~\ref{fig:basis-2d} also explains why latent-space updates are well suited to the EnKF for flows containing misaligned discontinuities. In the latent-EnKF, the analysis is formed by combining latent codes and then decoding the resulting code, so a shared feature represented by the decoder is retained as a single feature and evaluated at the location prescribed by the updated latent state. However, in the physical-space EnKF, Eqn.~\eqref{eq:affine} combines already decoded fields, so misaligned shocks and other coherent features are superposed directly in physical space, producing multiple spatially displaced copies of those features in the analysis state. The latent formulation therefore allows the EnKF analysis to move coherent features through the decoder while preserving their identity.

\section{Latent-EnKF for inference of geometry of an immersed body}
\label{sec:cylgeom}

In section~\ref{sec:representation} we showed that changes in the latent code move coherent flow features through the shared decoder. The shock-cylinder EnKF experiment of Sec.~\ref{sec:cylinder} considered uncertainty in the flow state while keeping the geometry fixed across the ensemble. Here, the geometry itself is uncertain. Each member carries a different elliptical body and a different shock Mach number. The Latent-EnKF must infer the unknown elliptical body's major- and minor-axis lengths, orientation, and position using measurements taken only in the surrounding flow.

\begin{figure}[!t]
\centering
\includegraphics[width=0.855\linewidth]{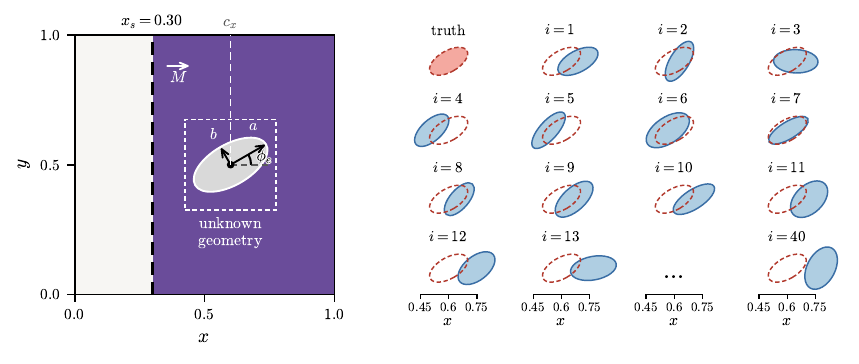}
\caption{Setup for the geometry-inference problem. Left: domain, shock location, Mach number, ellipse center, semi-axes and orientation. The control vector is $c=(r_a,r_b,\phi_e,c_x,M)$ with $r_a=2a/d$, $r_b=2b/d$, $d=0.30$ and $c_y=0.50$. Right: prior ensemble of immersed bodies at $t=0$, each drawn at its own center against the truth.}
\label{fig:cylgeom-setup}
\end{figure}

The flow configuration consists of a planar shock interacting with an elliptical body in $\Omega=[0,1]^2$ on a uniform $256\times256$ grid. The numerical method, including the sharp-interface immersed-boundary treatment, is the same as in Sec.~\ref{sec:experiments}. Each member is parameterized by $c=(a,b,\phi_e,c_x,M)$, where $a$ and $b$ are the ellipse semi-axes, $\phi_e$ is the ellipse orientation, $c_x$ is the center abscissa, and $M$ is the shock Mach number. The center ordinate is fixed at $c_y=0.5$, and the standoff distance between the shock and body center is fixed at $d=0.30$. The shock is initialized at $x_s=c_x-d$, so changes in body position also shift the initial shock location while preserving the same standoff distance. The truth controls are
\begin{equation*}
(r_a,r_b,\phi_e,c_x,M)_{\mathrm{truth}}
=
(0.75,0.375,30^\circ,0.60,2.0).
\end{equation*}
The uncertainties in the controls at $t=0$ are specified using normal distributions whose means are offset by $10\%$ from the true values:
\begin{equation*}
a\sim\mathcal{N}\bigl(0.1238,0.00875^{2}\bigr),
\quad
b\sim\mathcal{N}\bigl(0.0619,0.00875^{2}\bigr),
\quad
\phi_e\sim\mathcal{N}\bigl(33^\circ,(15^\circ)^{2}\bigr),
\end{equation*}
\begin{equation*}
c_x\sim\mathcal{N}\bigl(0.66,0.0667^{2}\bigr),
\quad
M\sim\mathcal{N}\bigl(1.90,0.03^{2}\bigr),
\end{equation*}
Figure~\ref{fig:cylgeom-setup} shows both the geometric parameterization and a few samples of the ensemble geometries initialized at $t=0$.

The observations follow Sec.~\ref{sec:cylinder}. Time-cumulative $|\nabla \rho|$ over each assimilation interval is sampled at eight equally spaced times and projected onto a $32\times32$ sensor grid. A fixed square mask of side $0.25$, centered at $(0.60,0.50)$, removes measurements over the body and leaves $960$ observed cells. The observation-error model uses a relative coefficient of $0.12$ with additive floor $0.06$. The ensemble contains $n_e=40$ members, the latent dimension is $n_z=16$, and the decoder is retrained on the forecast ensemble at every assimilation cycle. After a flow transient that brings the shock into interaction with the body, five analyses are performed at $t=\{0.13, 0.165, 0.20, 0.23, 0.27\}$.

Three modifications extend Algorithm~\ref{alg:fs-enkf} to uncertain immersed geometry. Each member first carries its own ellipse parameters into the flow solver, so its immersed-boundary mask, ghost cells and wall normals evolve with that member. The decoder is trained over the full computational domain, including both fluid and solid cells, with the primitive variables in the solid region set to zero. This allows the latent representation to encode the body together with the surrounding flow. After the analysis step on the latent codes, the geometry is updated by fitting an ellipse to the zero-density region of each decoded analysis field. Any cells uncovered by the moving body that contain a zero state are initialized from neighboring fluid values using the fresh-cell procedure of Udaykumar et al.~\cite{udaykumar2001sharp} before the next forecast step. These three modifications allow the latent analysis to modify the geometry while maintaining a valid state for the immersed-boundary solver.

\begin{figure}[t!]
\centering
\includegraphics[width=\linewidth]{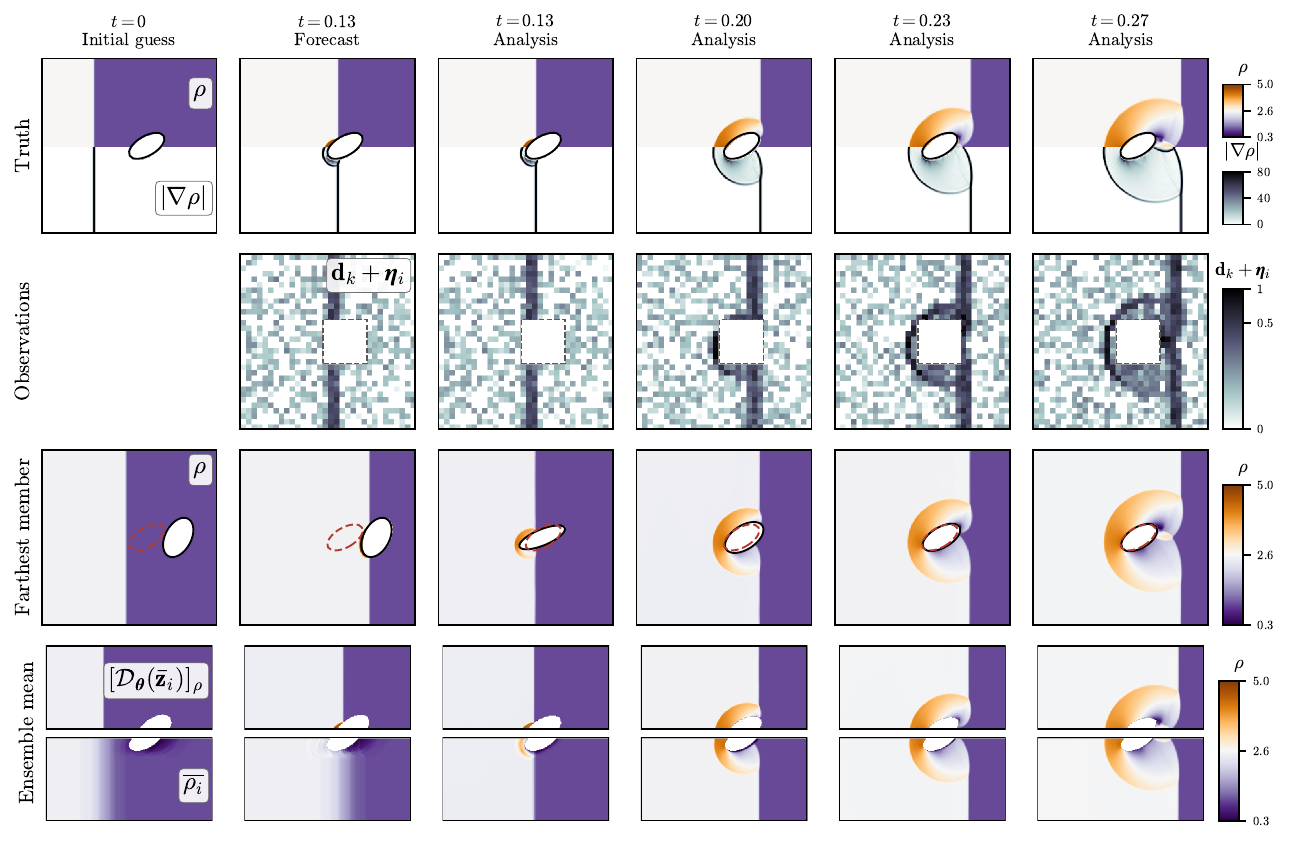}
\caption{\textbf{Geometry inference with the latent-EnKF.} Columns show the initial condition, forecast and successive analyses. Row 1: truth, with density $\rho$ and instantaneous $|\nabla\rho|$. Row 2: noisy cumulative $|\nabla\rho|$ observations outside the fixed body mask. Row 3: density of the ensemble member farthest from the mean together with its ellipse estimate. Row 4: mean obtained by decoding the mean latent code compared with the mean formed directly in physical space.}
\label{fig:cylgeom-assim}
\end{figure}

\begin{figure}[!t]
\centering
\includegraphics[width=0.85\textwidth]{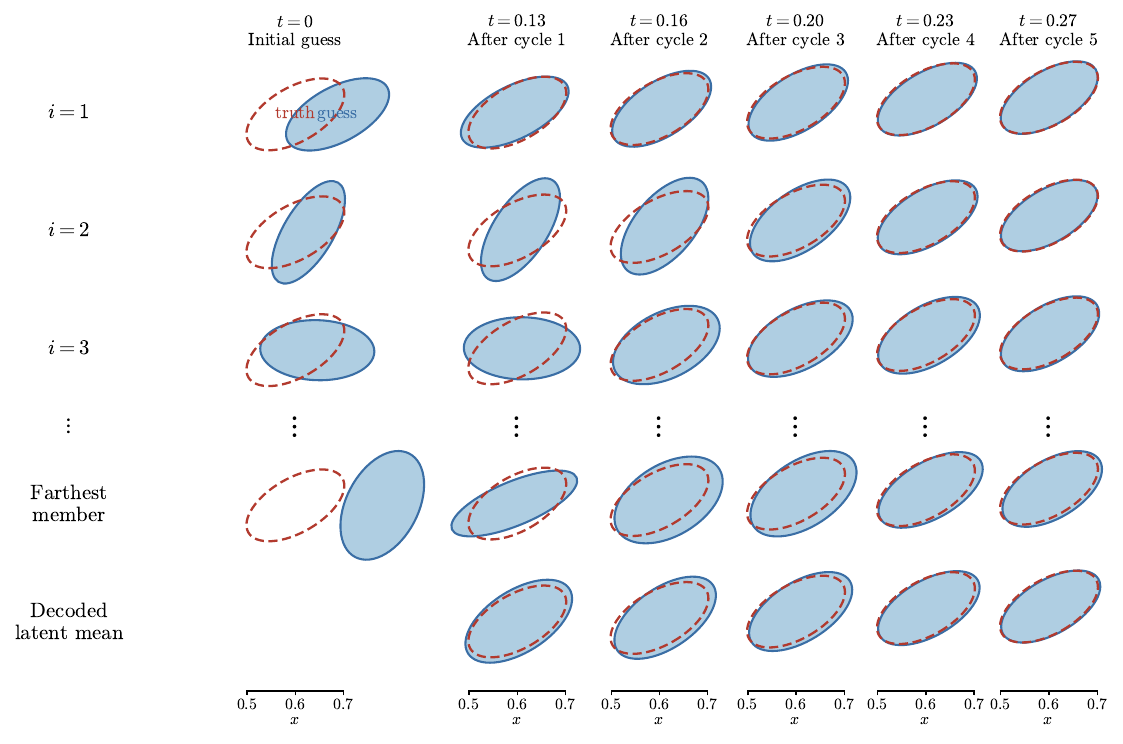}
\caption{Convergence of the inferred body geometry. Rows show selected ensemble members, including the member with the largest normalized geometry error. Columns show the prior and the five successive analyses. Each panel compares the member estimate with the truth.}
\label{fig:cylgeom-shape}
\end{figure}

\begin{figure}[!t]
\centering
\includegraphics[width=0.925\linewidth]{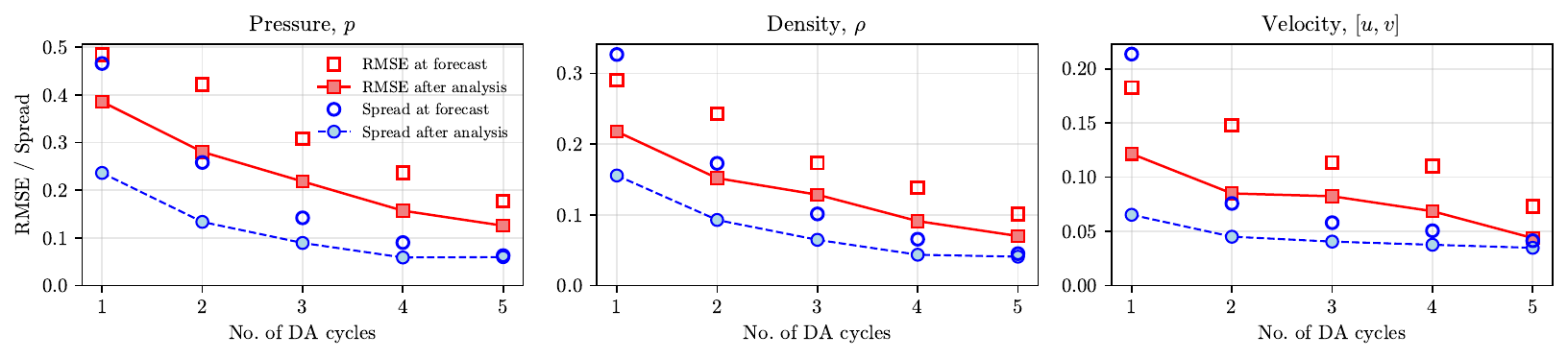}
\caption{Per-cycle RMSE and ensemble spread for pressure $p$, density $\rho$ and velocity magnitude $[u,v]$, shown for the forecast and analysis at each assimilation cycle.}
\label{fig:cylgeom-rmse}
\end{figure}

Figure~\ref{fig:cylgeom-assim} shows that the analysis corrects the geometry and the surrounding flow simultaneously. The farthest member at $t=0$ carries a body that differs substantially from the truth in position, orientation and aspect ratio. After the first analysis, the body of that member is already displaced toward the true location, and the subsequent cycles progressively align both the geometry and the associated shock interaction. By the final cycle, the reflected and transmitted flow features around the inferred body closely follow those of the truth. The mean obtained by decoding the mean latent code retains a distinct localized flow features, whereas direct averaging in physical space diffuses features whose locations vary across the ensemble.

The convergence of the body parameters is isolated in Fig.~\ref{fig:cylgeom-shape}. Comparing the forecast ellipse parameters between the initial guess and the final analysis, the ensemble RMSE decreases from $19.89$ to $0.98$ grid cells in $c_x$, from $15.1^\circ$ to $1.01^\circ$ in $\phi_e$, from $3.11$ to $0.48$ cells in the major semi-axis, and from $2.64$ to $0.36$ cells in the minor semi-axis. Thus, all three inferred length scales reach approximately one-grid-cell accuracy by the final cycle. The figure also shows that even the member farthest from the true geometry at $t=0$ converges within five analysis cycles. The ensemble spread contracts over the same interval to values comparable with the corresponding RMSE, indicating that the reduction in geometric uncertainty accompanies the reduction in estimation error.

The improvement in geometry is accompanied by a corresponding improvement in the flow estimate. Figure~\ref{fig:cylgeom-rmse} shows that the largest correction occurs during the early cycles, when the geometric spread is also largest. The error does not decrease monotonically at every individual cycle, particularly for pressure and density, but the forecast and analysis statistics exhibit a clear overall contraction across the five cycles. The ensemble spread follows the same trend and remains of comparable magnitude to the RMSE by the final analysis. 

\section{Conclusion}
\label{sec:conclusions}

A feature-preserving latent-space EnKF has been developed for data assimilation of shock-laden compressible flows. The approach addresses a central limitation of the standard EnKF: when discontinuities are misaligned across the forecast ensemble, the linear analysis update can generate non-monotone states and spurious oscillations near the discontinuity. Three aspects of the EnKF update explain this failure. (i) Misaligned discontinuities produce bimodal pointwise ensemble statistics by mixing the left and right states across the discontinuity region, violating the Gaussian assumptions of the EnKF. However, bimodality alone is not sufficient to produce non-monotone analysis states. (ii) Expressing the analysis state as a linear combination of the forecast members shows that negative weights can get assigned to some of the forecast ensemble members and reverse the direction of jumps in the weighted forecast states. These reversed jumps can cause the analysis state to become non-monotone and oscillatory. (iii) The same failure appears in the Kalman-gain view. To translate a discontinuity coherently, the gain profile must have a step-like form. Instead, the gain profiles of physical-space EnKF yield spread-out non-step like correction profiles over the region occupied by the forecast fronts. Such corrections act alongside the displaced forecast discontinuity and can produce overshoots and undershoots whose magnitude scales with the difference between the mean states on either side of the discontinuity, leading to a non-monotone analysis state.

The proposed latent-EnKF transfers the linear analysis from physical space to a latent space learned through a CC-auto-decoder, where variations in shock location, contact discontinuities, slip lines, and other flow features are encoded as smooth variations in the latent coordinates. Numerical experiments on a Sod shock tube and a Mach~2 shock interaction with a cylinder demonstrated stable and accurate assimilation from sparse and noisy observations, including schlieren-like cumulative density-gradient measurements, while preserving sharp flow structures throughout the assimilation cycles. The latent-EnKF was further extended to infer the geometry of an unknown immersed body from noisy schlieren measurements of the surrounding flow, with the body position, size, and orientation recovered through latent-EnKF.


The accuracy and feature-preserving behavior of the latent-EnKF can be traced to five properties of the learned representation. (i) The latent coordinates retain near-Gaussian ensemble statistics, making the CC-auto-decoder's latent space better suited to the linear EnKF update. (ii) Linear superposition in physical space combines spatially displaced discontinuities and can produce nonphysical intermediate states, whereas the same operation in latent space produces a single coherent realization in which the flow features move continuously between ensemble members. (iii) After decoding, the latent-space Kalman update produces effective gain profiles with step-like form near discontinuities, allowing the fronts and plateau states to be corrected without introducing the non-monotonicity associated with the distributed kernels of the physical-space EnKF. (iv) The active latent directions encode the dominant physical degrees of freedom of the ensemble, including variations in shock and body location. (v) The latent code controls the locations and amplitudes of the flow features by progressively morphing the internal coordinates through the layers of the decoder. The shared downstream decoder then maps these morphed coordinates back to the physical field, so changes in the latent code move and modify common flow features coherently rather than creating separate member-specific representations. These properties allow the linear Kalman update to modify the feature variations of the ensemble while preserving coherent physical states after decoding.
The proposed latent-EnKF is solver-agnostic and extends naturally to three-dimensional configurations and other physical systems featuring discontinuities, such as detonation fronts, material interfaces, and phase boundaries. 

\section*{Acknowledgments} This work was supported by the AFOSR\slash AFRL Center of Excellence in Assimilation of Flow Features in Compressible Reacting Flows under award number FA9550-25-1-0011. \par

\section*{Code availability} Data and codes are available upon request.

\appendix

\section{The entries of Kalman gain as regression slopes}
\label{app:gain-regression}

This appendix algebraically simplifies each entry of the Kalman gain and shows how each entry can be interpreted as the slope of the best-fit line through the ensemble cloud in the sensor-state plane (Fig.~\ref{fig:gain-scatter}(a2,b2)). Starting from the Kalman gain,
\begin{equation}
\mathbf{K}
=
\mathbf{Q}\mathbf{Y}^{\!\top}
\bigl(\mathbf{Y}\mathbf{Y}^{\!\top}+\mathbf{R}\bigr)^{-1},
\end{equation}
define
\begin{equation}
\mathbf{C}_{qy}
=
\mathbf{Q}\mathbf{Y}^{\!\top},
\qquad
\mathbf{C}_{yy}
=
\mathbf{Y}\mathbf{Y}^{\!\top}+\mathbf{R},
\end{equation}
so that
\begin{equation}
\mathbf{K}
=
\mathbf{C}_{qy}\mathbf{C}_{yy}^{-1}
=
\bigl[\mathbf{k}_1,\ldots,\mathbf{k}_{n_y}\bigr].
\end{equation}
Using the perturbation matrices of Eqn.~\eqref{eq:perturb}, the entries of these covariance matrices are
\begin{equation}
(C_{qy})_{j\ell}
=
\frac{1}{n_e-1}
\sum_{i=1}^{n_e}
\bigl(q_{ij}^{f}-\bar q_j^{f}\bigr)
\bigl(y_{i\ell}^{f}-\bar y_{\ell}^{f}\bigr),
\end{equation}
and
\begin{equation}
(C_{yy})_{m\ell}
=
\frac{1}{n_e-1}
\sum_{i=1}^{n_e}
\bigl(y_{im}^{f}-\bar y_m^{f}\bigr)
\bigl(y_{i\ell}^{f}-\bar y_{\ell}^{f}\bigr)
+
R_{m\ell}.
\end{equation}
Hence, the entry of the gain profile $\mathbf{k}_{\ell}$ at grid point $x_j$ is
\begin{equation}
k_{\ell}(x_j)
=
K_{j\ell}
=
\sum_{m=1}^{n_y}
(C_{qy})_{jm}
(C_{yy}^{-1})_{m\ell}.
\end{equation}
For a single sensor, this reduces to
\begin{align}
k_{\ell}(x_j)
&=
\frac{
\displaystyle
\frac{1}{n_e-1}
\sum_{i=1}^{n_e}
\bigl(q_{ij}^{f}-\bar q_j^{f}\bigr)
\bigl(y_{i\ell}^{f}-\bar y_{\ell}^{f}\bigr)
}{
\displaystyle
\frac{1}{n_e-1}
\sum_{i=1}^{n_e}
\bigl(y_{i\ell}^{f}-\bar y_{\ell}^{f}\bigr)^2
+
R_{\ell\ell}
}
\nonumber\\
&=
\frac{
\operatorname{Cov}(q_j^{f},y_{\ell}^{f})
}{
\operatorname{Var}(y_{\ell}^{f})+R_{\ell\ell}
}.
\label{eq:app-slope}
\end{align}
Let $\widetilde y_{\ell}^{f}=y_{\ell}^{f}+\eta_{\ell}$ denote the noisy sensor measurement of Sec.~\ref{sec:gain}, where $\eta_{\ell}$ has zero mean, variance $R_{\ell\ell}$, and is independent of $q_j^{f}$ and $y_{\ell}^{f}$. For the numerator,
\begin{equation}
\begin{aligned}
\operatorname{Cov}\!\left(q_j^{f},\widetilde y_{\ell}^{f}\right)
&=
\operatorname{Cov}\!\left(q_j^{f},y_{\ell}^{f}+\eta_{\ell}\right)\\
&=
\operatorname{Cov}\!\left(q_j^{f},y_{\ell}^{f}\right)
+
\operatorname{Cov}\!\left(q_j^{f},\eta_{\ell}\right)\\
&=
\operatorname{Cov}\!\left(q_j^{f},y_{\ell}^{f}\right),
\end{aligned}
\end{equation}
since $\eta_{\ell}$ is independent of $q_j^{f}$. For the denominator,
\begin{equation}
\begin{aligned}
\operatorname{Var}\!\left(\widetilde y_{\ell}^{f}\right)
&=
\operatorname{Var}\!\left(y_{\ell}^{f}+\eta_{\ell}\right)\\
&=
\operatorname{Var}\!\left(y_{\ell}^{f}\right)
+
\operatorname{Var}\!\left(\eta_{\ell}\right)
+
2\operatorname{Cov}\!\left(y_{\ell}^{f},\eta_{\ell}\right)\\
&=
\operatorname{Var}\!\left(y_{\ell}^{f}\right)
+
R_{\ell\ell},
\end{aligned}
\end{equation}
because $\eta_{\ell}$ is independent of $y_{\ell}^{f}$ and
$\operatorname{Var}(\eta_{\ell})=R_{\ell\ell}$. Therefore,
\begin{equation}
k_{\ell}(x_j)
=
\frac{
\operatorname{Cov}\!\left(q_j^{f},\widetilde y_{\ell}^{f}\right)
}{
\operatorname{Var}\!\left(\widetilde y_{\ell}^{f}\right)
}.
\end{equation}
Thus, $k_{\ell}(x_j)$ is the slope of the fitted line through the forecast ensemble points
$[\widetilde y_{\ell}^{f},q_j^{f}]$ in the sensor-state plane, as illustrated in Fig.~\ref{fig:gain-scatter}(a2,b2).

\section{Time-series latent-walk between two snapshots}
\label{app:riemann-walk}

\begin{figure}[!t]
\centering
\includegraphics[width=0.90\textwidth]{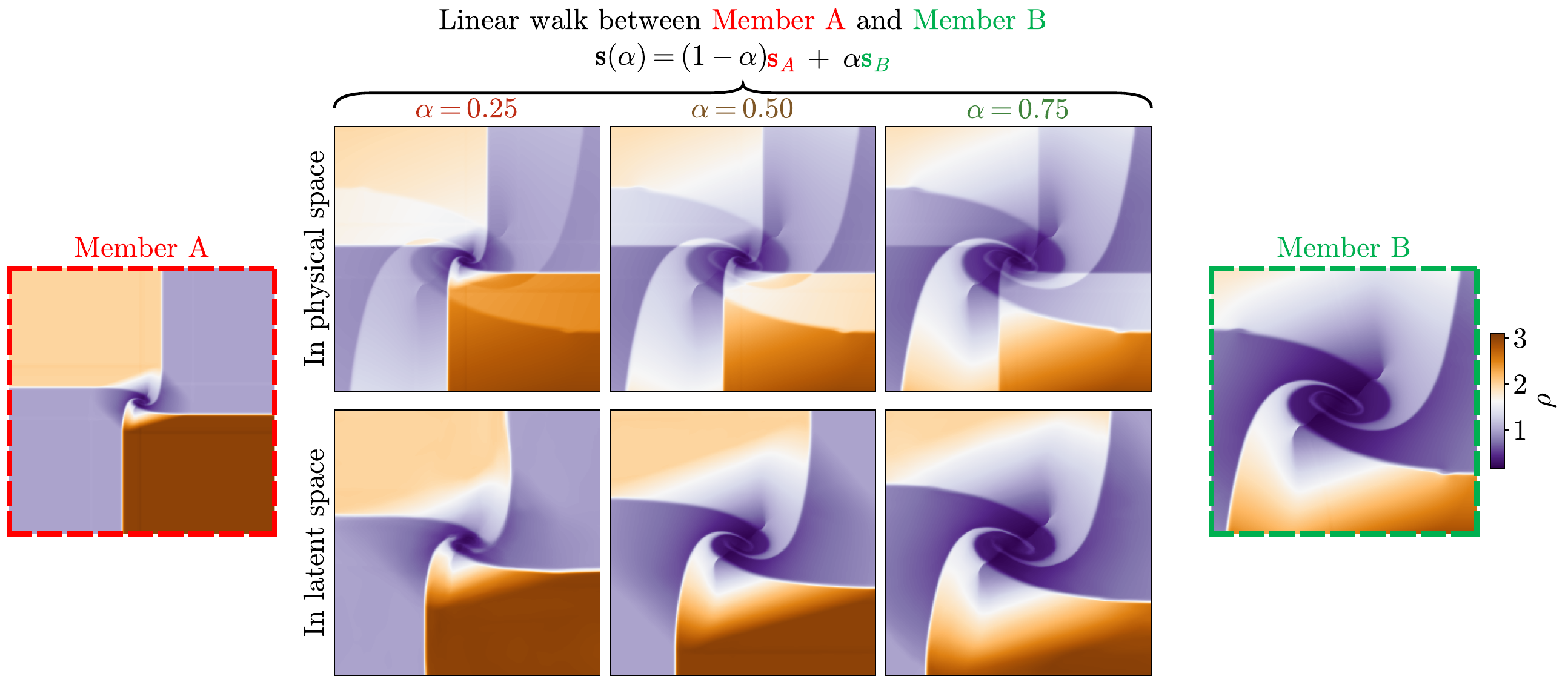}
\caption{Density $\rho$ for configuration 6 of the two-dimensional Riemann problem. Members $A$ and $B$ correspond to snapshots at $t=0.1$ and $t=0.6$. Intermediate states at $\alpha=0.25$, $0.50$, and $0.75$ are formed by linear superposition in physical space and by linear superposition in the latent space of a CC-auto-decoder fitted to snapshots at $t=0.1$, $0.35$, and $0.6$.}
\label{fig:superpose-riemann}
\end{figure}

The latent superposition test of Sec.~\ref{sec:superpose} is extended here from two ensemble members at a fixed time to two states at different times along a nonlinear trajectory governed by the Euler equations. Members $A$ and $B$ are taken from configuration 6 of the two-dimensional Riemann problem \cite{lax1998solution} at $t=0.1$ and $t=0.6$. The solution contains interacting contacts and shear layers that develop into a spiral structure. A CC-auto-decoder with $n_z=16$ is fitted to the snapshots at $t=0.1$, $0.35$, and $0.6$. Linear superposition between the latent codes of $A$ and $B$ is then evaluated at $\alpha=0.25$, $0.50$, and $0.75$, corresponding to reference times $t=0.22$, $0.35$, and $0.47$. Figure~\ref{fig:superpose-riemann} compares these latent superpositions with linear superposition in physical space. Physical-space superposition overlays the contacts and spiral structures from the two endpoint states, whereas latent superposition produces a single set of contacts and shear layers that evolves continuously from $A$ to $B$.

The straight-line path between $\mathbf{z}_A$ and $\mathbf{z}_B$ therefore decodes to states that remain close to the physical trajectory between the two snapshots. This behavior is relevant to reduced-order models that evolve the state directly in latent space, as demonstrated with autoencoders \cite{lee2020model,champion2019data,lusch2018deep,maulik2021reduced,fresca2021comprehensive,vlachas2022multiscale,wiewel2019latent}, auto-decoders and neural fields \cite{chen2023crom,pan2023neural,yin2023continuous,serrano2023operator}, and latent-space data assimilation \cite{peyron2021latent,amendola2021data,cheng2023machine}. In the present latent-EnKF, the numerical solver advances the forecast in physical space and the latent representation is used only for the analysis update. The results indicate that interpolation in latent-space remains effective for states separated in time and may therefore enable the flow dynamics to be evolved in a low-dimensional latent space, reducing the computational cost relative to advancing the full high-dimensional physical state.

\bibliographystyle{elsarticle-num}
\bibliography{refs}

@book{asch2016,
  author    = {Asch, Marc and Bocquet, Marc and Nodet, Maelle},
  title     = {Data Assimilation: {M}ethods, Algorithms, and Applications},
  publisher = {SIAM},
  year      = {2016},
}

@incollection{zaki2025data,
  author    = {Zaki, Tamer A. and Wang, Mengze},
  title     = {Data assimilation and flow estimation},
  booktitle = {Data Driven Analysis and Modeling of Turbulent Flows},
  publisher = {Elsevier},
  pages     = {129--181},
  year      = {2025},
}

@article{buchta2021observation,
  title={Observation-infused simulations of high-speed boundary-layer transition},
  author={Buchta, David A and Zaki, Tamer A},
  journal={Journal of Fluid Mechanics},
  volume={916},
  pages={A44},
  year={2021},
  publisher={Cambridge University Press}
}

@article{buchta2022assimilation,
  title={Assimilation of wall-pressure measurements in high-speed flow over a cone},
  author={Buchta, David A and Laurence, Stuart J and Zaki, Tamer A},
  journal={Journal of Fluid Mechanics},
  volume={947},
  pages={R2},
  year={2022},
  publisher={Cambridge University Press}
}

@article{morra2024ml,
  title={{ML} for fast assimilation of wall-pressure measurements from hypersonic flow over a cone},
  author={Morra, Pierluigi and Meneveau, Charles and Zaki, Tamer A},
  journal={Scientific Reports},
  volume={14},
  number={1},
  pages={12853},
  year={2024},
  publisher={Nature Publishing Group UK London}
}

@article{wang2025domain,
  title={Domain of dependence for wall-pressure measurements in high-speed boundary layers},
  author={Wang, Qi and Zaki, Tamer A},
  journal={Journal of Fluid Mechanics},
  volume={1009},
  pages={A67},
  year={2025},
  publisher={Cambridge University Press}
}

@article{zaki2025turbulence,
   author = "Zaki, Tamer A.",
   title = "Turbulence from an Observer Perspective", 
   journal= "Annual Review of Fluid Mechanics",
   year = "2025",
   volume = "57",
   number = "Volume 57, 2025",
   pages = "311-334",
   publisher = "Annual Reviews",
   issn = "1545-4479",
   type = "Journal Article"
}

@inproceedings{houba2024,
  author    = {Houba, T. and Edoh, A. and Munipalli, R. and Harvazinski, M. E.},
  title     = {Sequential data assimilation in flows with shocks},
  booktitle = {AIAA SCITECH 2024 Forum},
  year      = {2024},
  note      = {Paper 0587},
}

@article{morzfeld2019,
  author  = {Morzfeld, M. and Hodyss, D.},
  title   = {Gaussian approximations in filters and smoothers for data assimilation},
  journal = {Tellus A: Dynamic Meteorology and Oceanography},
  volume  = {71},
  pages   = {1600344},
  year    = {2019},
}

@article{zhou2026neural,
  title={Neural ensemble Kalman filter: Data assimilation for compressible flows with shocks},
  author={Zhou, Xu-Hui and Beronilla, Lorenzo and Sleeman, Michael K and Hu, Hangchuan and Morzfeld, Matthias and Stuart, Andrew M and Zaki, Tamer A},
  journal={arXiv preprint arXiv:2602.23461},
  year={2026}
}

@inproceedings{edoh2025,
  author    = {Edoh, A. and Houba, T. and Munipalli, R. and Harvazinski, M. E.},
  title     = {Sequential ensemble {K}alman filtering of compressible flows with shocks: enforcing positivity},
  booktitle = {AIAA SCITECH 2025 Forum},
  year      = {2025},
  note      = {Paper 0918},
}

@inproceedings{hansen2024,
  author    = {Hansen, J. J. and Brouzet, D. and Ihme, M.},
  title     = {A normal-score ensemble {K}alman filter for {1D} shock waves},
  booktitle = {AIAA SCITECH 2024 Forum},
  year      = {2024},
  note      = {Paper 1022},
}

@article{zhou2011,
  author  = {Zhou, H. and Gomez-Hernandez, J. J. and Franssen, H.-J. H. and Li, L.},
  title   = {An approach to handling non-{G}aussianity of parameters and state variables in ensemble {K}alman filtering},
  journal = {Advances in Water Resources},
  volume  = {34},
  pages   = {844--864},
  year    = {2011},
}

@article{li2024,
  author  = {Li, T. and Gelb, A. and Lee, Y.},
  title   = {A structurally informed data assimilation approach for nonlinear partial differential equations},
  journal = {Journal of Computational Physics},
  volume  = {519},
  pages   = {113396},
  year    = {2024},
}

@article{chaudhuri2011use,
  author  = {Chaudhuri, A. and Hadjadj, A. and Chinnayya, A.},
  title   = {On the use of immersed boundary methods for shock/obstacle interactions},
  journal = {Journal of Computational Physics},
  volume  = {230},
  number  = {5},
  pages   = {1731--1748},
  year    = {2011},
}

@inproceedings{ke2026observability,
  author    = {Ke, G. and Grauer, S. J. and Zaki, T. A.},
  title     = {Observability of initial states in one-dimensional inviscid flows with shocks},
  booktitle = {AIAA SCITECH 2026 Forum},
  year      = {2026},
  note      = {Paper 0493},
}

@inproceedings{park2019deepsdf,
  author    = {Park, J. J. and Florence, P. and Straub, J. and Newcombe, R. and Lovegrove, S.},
  title     = {{DeepSDF}: learning continuous signed distance functions for shape representation},
  booktitle = {Proc. IEEE/CVF Conf. on Computer Vision and Pattern Recognition (CVPR)},
  pages     = {165--174},
  year      = {2019},
}

@article{tong2026latent,
  author  = {Tong, Xin T. and Wang, Yanyan and Yan, Liang},
  title   = {Latent autoencoder ensemble {K}alman filter for nonlinear data assimilation},
  journal = {arXiv preprint arXiv:2603.06752},
  year    = {2026},
}

@article{lax1998solution,
  title={Solution of two-dimensional Riemann problems of gas dynamics by positive schemes},
  author={Lax, Peter D and Liu, Xu-Dong},
  journal={SIAM Journal on Scientific Computing},
  volume={19},
  number={2},
  pages={319--340},
  year={1998},
  publisher={SIAM}
}

@article{fan2026physically,
  title={Physically consistent global atmospheric data assimilation with machine learning in latent space},
  author={Fan, Hang and Bai, Lei and Fei, Ben and Xiao, Yi and Chen, Kun and Liu, Yubao and Qu, Yongquan and Ling, Fenghua and Gentine, Pierre},
  journal={Science Advances},
  volume={12},
  number={1},
  pages={eaea4248},
  year={2026},
  publisher={American Association for the Advancement of Science}
}

@article{evensen2003ensemble,
  author  = {Evensen, Geir},
  title   = {The ensemble {K}alman filter: theoretical formulation and practical implementation},
  journal = {Ocean Dynamics},
  volume  = {53},
  number  = {4},
  pages   = {343--367},
  year    = {2003},
}

@book{evensen2009data,
  title={Data assimilation: the ensemble Kalman filter},
  author={Evensen, Geir},
  year={2009},
  publisher={Springer}
}

@article{li2025structurally,
  title={Structurally informed data assimilation in two dimensions},
  author={Li, Tongtong and Gelb, Anne and Lee, Yoonsang},
  journal={arXiv preprint arXiv:2510.06369},
  year={2025}
}

@article{jiang1996efficient,
  title={Efficient implementation of weighted ENO schemes},
  author={Jiang, Guang-Shan and Shu, Chi-Wang},
  journal={Journal of computational physics},
  volume={126},
  number={1},
  pages={202--228},
  year={1996},
  publisher={Elsevier}
}

@article{toro1994restoration,
  title={Restoration of the contact surface in the HLL-Riemann solver},
  author={Toro, Eleuterio F and Spruce, Michael and Speares, William},
  journal={Shock waves},
  volume={4},
  number={1},
  pages={25--34},
  year={1994},
  publisher={Springer}
}

@article{chandravamsi2023application,
  title={On the application of gradient based reconstruction for flow simulations on generalized curvilinear and dynamic mesh domains},
  author={Chandravamsi, Hemanth and Chamarthi, Amareshwara Sainadh and Hoffmann, Natan and Frankel, Steven H},
  journal={Computers \& Fluids},
  volume={258},
  pages={105859},
  year={2023},
  publisher={Elsevier}
}

@article{gottlieb2001strong,
  title={Strong stability-preserving high-order time discretization methods},
  author={Gottlieb, Sigal and Shu, Chi-Wang and Tadmor, Eitan},
  journal={SIAM review},
  volume={43},
  number={1},
  pages={89--112},
  year={2001},
  publisher={SIAM}
}

@article{fan2025novel,
  title={A novel latent space data assimilation framework with autoencoder-observation to latent space (AE-O2L) network. Part I: The observation-only analysis method},
  author={Fan, Hang and Liu, Yubao and Huo, Zhaoyang and Liu, Yuewei and Shi, Yueqin and Li, Yang},
  journal={Monthly Weather Review},
  volume={153},
  number={8},
  pages={1335--1348},
  year={2025},
  publisher={American Meteorological Society}
}

@article{pasmans2026ensemble,
  title={Ensemble Kalman filter in latent space using a variational autoencoder pair},
  author={Pasmans, Ivo and Chen, Yumeng and Sebastian Finn, Tobias and Bocquet, Marc and Carrassi, Alberto},
  journal={Quarterly Journal of the Royal Meteorological Society},
  volume={152},
  number={775},
  pages={e70070},
  year={2026},
  publisher={Wiley Online Library}
}

@article{chen2023reduced,
  title={Reduced-order autodifferentiable ensemble Kalman filters},
  author={Chen, Yuming and Sanz-Alonso, Daniel and Willett, Rebecca},
  journal={Inverse Problems},
  volume={39},
  number={12},
  pages={124001},
  year={2023},
  publisher={IOP Publishing}
}

@article{snyder2008obstacles,
  author  = {Snyder, Chris and Bengtsson, Thomas and Bickel, Peter and Anderson, Jeff},
  title   = {Obstacles to high-dimensional particle filtering},
  journal = {Monthly Weather Review},
  volume  = {136},
  number  = {12},
  pages   = {4629--4640},
  year    = {2008},
}

@article{poterjoy2016localized,
  author  = {Poterjoy, Jonathan},
  title   = {A localized particle filter for high-dimensional nonlinear systems},
  journal = {Monthly Weather Review},
  volume  = {144},
  number  = {1},
  pages   = {59--76},
  year    = {2016},
}

@article{bengtsson2003toward,
  author  = {Bengtsson, Thomas and Snyder, Chris and Nychka, Doug},
  title   = {Toward a nonlinear ensemble filter for high-dimensional systems},
  journal = {Journal of Geophysical Research: Atmospheres},
  volume  = {108},
  number  = {D24},
  pages   = {8775},
  year    = {2003},
}

@article{hoteit2008new,
  author  = {Hoteit, Ibrahim and Pham, Dinh-Tuan and Triantafyllou, George and Korres, Gerasimos},
  title   = {A new approximate solution of the optimal nonlinear filter for data assimilation in meteorology and oceanography},
  journal = {Monthly Weather Review},
  volume  = {136},
  number  = {1},
  pages   = {317--334},
  year    = {2008},
}

@article{grooms2022comparison,
  author  = {Grooms, Ian},
  title   = {A comparison of nonlinear extensions to the ensemble {K}alman filter: {G}aussian anamorphosis and two-step ensemble filters},
  journal = {Computational Geosciences},
  volume  = {26},
  pages   = {633--650},
  year    = {2022},
}

@article{anderson2010non,
  author  = {Anderson, Jeffrey L.},
  title   = {A non-{G}aussian ensemble filter update for data assimilation},
  journal = {Monthly Weather Review},
  volume  = {138},
  number  = {11},
  pages   = {4186--4198},
  year    = {2010},
}

@article{bertino2003sequential,
  author  = {Bertino, Laurent and Evensen, Geir and Wackernagel, Hans},
  title   = {Sequential data assimilation techniques in oceanography},
  journal = {International Statistical Review},
  volume  = {71},
  number  = {2},
  pages   = {223--241},
  year    = {2003},
}

@article{simon2009application,
  author  = {Simon, Ehouarn and Bertino, Laurent},
  title   = {Application of the {G}aussian anamorphosis to assimilation in a {3-D} coupled physical-ecosystem model of the {N}orth {A}tlantic with the {EnKF}: a twin experiment},
  journal = {Ocean Science},
  volume  = {5},
  number  = {4},
  pages   = {495--510},
  year    = {2009},
}

@article{bocquet2014iterative,
  author  = {Bocquet, Marc and Sakov, Pavel},
  title   = {An iterative ensemble {K}alman smoother},
  journal = {Quarterly Journal of the Royal Meteorological Society},
  volume  = {140},
  number  = {682},
  pages   = {1521--1535},
  year    = {2014},
}

@article{spantini2022coupling,
  author  = {Spantini, Alessio and Baptista, Ricardo and Marzouk, Youssef},
  title   = {Coupling techniques for nonlinear ensemble filtering},
  journal = {SIAM Review},
  volume  = {64},
  number  = {4},
  pages   = {921--953},
  year    = {2022},
}

@article{ramgraber2023ensemble,
  author  = {Ramgraber, Maximilian and Baptista, Ricardo and McLaughlin, Dennis and Marzouk, Youssef},
  title   = {Ensemble transport smoothing. {P}art {I}: {U}nified framework},
  journal = {Journal of Computational Physics: X},
  volume  = {17},
  pages   = {100134},
  year    = {2023},
}

@inproceedings{rozet2023score,
  author    = {Rozet, Fran\c{c}ois and Louppe, Gilles},
  title     = {Score-based data assimilation},
  booktitle = {Advances in Neural Information Processing Systems},
  volume    = {36},
  year      = {2023},
}

@article{ravela2007data,
  author  = {Ravela, Sai and Emanuel, Kerry and McLaughlin, Dennis},
  title   = {Data assimilation by field alignment},
  journal = {Physica D: Nonlinear Phenomena},
  volume  = {230},
  number  = {1--2},
  pages   = {127--145},
  year    = {2007},
}

@article{beezley2008morphing,
  author  = {Beezley, Jonathan D. and Mandel, Jan},
  title   = {Morphing ensemble {K}alman filters},
  journal = {Tellus A: Dynamic Meteorology and Oceanography},
  volume  = {60},
  number  = {1},
  pages   = {131--140},
  year    = {2008},
}

@article{nehrkorn2015correcting,
  author  = {Nehrkorn, Thomas and Woods, Bryan K. and Hoffman, Ross N. and Auligné, Thomas},
  title   = {Correcting for position errors in variational data assimilation},
  journal = {Monthly Weather Review},
  volume  = {143},
  number  = {4},
  pages   = {1368--1381},
  year    = {2015},
}

@article{feyeux2018optimal,
  author  = {Feyeux, Nelson and Vidard, Arthur and Nodet, Ma\"elle},
  title   = {Optimal transport for variational data assimilation},
  journal = {Nonlinear Processes in Geophysics},
  volume  = {25},
  number  = {1},
  pages   = {55--66},
  year    = {2018},
}

@article{tamang2021ensemble,
  author  = {Tamang, Sagar K. and Ebtehaj, Ardeshir and van Leeuwen, Peter J. and Zou, Dongmian and Lerman, Gilad},
  title   = {Ensemble {R}iemannian data assimilation over the {W}asserstein space},
  journal = {Nonlinear Processes in Geophysics},
  volume  = {28},
  number  = {3},
  pages   = {295--309},
  year    = {2021},
}

@inproceedings{chen2023crom,
  author    = {Chen, Peter Yichen and Xiang, Jinxu and Cho, Dong Heon and Chang, Yue and Pershing, G. A. and Maia, Henrique Teles and Chiaramonte, Maurizio M. and Carlberg, Kevin T. and Grinspun, Eitan},
  title     = {{CROM}: {C}ontinuous reduced-order modeling of {PDE}s using implicit neural representations},
  booktitle = {International Conference on Learning Representations (ICLR)},
  year      = {2023},
}

@article{pan2023neural,
  author  = {Pan, Shaowu and Brunton, Steven L. and Kutz, J. Nathan},
  title   = {Neural implicit flow: a mesh-agnostic dimensionality reduction paradigm of spatio-temporal data},
  journal = {Journal of Machine Learning Research},
  volume  = {24},
  number  = {41},
  pages   = {1--60},
  year    = {2023},
}

@inproceedings{yin2023continuous,
  author    = {Yin, Yuan and Kirchmeyer, Matthieu and Franceschi, Jean-Yves and Rakotomamonjy, Alain and Gallinari, Patrick},
  title     = {Continuous {PDE} dynamics forecasting with implicit neural representations},
  booktitle = {International Conference on Learning Representations (ICLR)},
  year      = {2023},
}

@inproceedings{serrano2023operator,
  author    = {Serrano, Louis and Le Boudec, Lise and Kassa\"{\i} Koupa\"{\i}, Armand and Wang, Thomas X. and Yin, Yuan and Vittaut, Jean-No\"el and Gallinari, Patrick},
  title     = {Operator learning with neural fields: {T}ackling {PDE}s on general geometries},
  booktitle = {Advances in Neural Information Processing Systems},
  volume    = {36},
  year      = {2023},
}

@inproceedings{amendola2021data,
  author    = {Amendola, Maddalena and Arcucci, Rossella and Mottet, Laetitia and Casas, C\'esar Quilodr\'an and Fan, Shiwei and Pain, Christopher and Linden, Paul and Guo, Yi-Ke},
  title     = {Data assimilation in the latent space of a convolutional autoencoder},
  booktitle = {Computational Science -- ICCS 2021},
  series    = {Lecture Notes in Computer Science},
  volume    = {12746},
  pages     = {373--386},
  publisher = {Springer},
  year      = {2021},
}

@article{peyron2021latent,
  author  = {Peyron, Mathis and Fillion, Anthony and G\"urol, Selime and Marchais, Victor and Gratton, Serge and Boudier, Pierre and Goret, Gael},
  title   = {Latent space data assimilation by using deep learning},
  journal = {Quarterly Journal of the Royal Meteorological Society},
  volume  = {147},
  number  = {740},
  pages   = {3759--3777},
  year    = {2021},
}

@article{cheng2023machine,
  author  = {Cheng, Sibo and Quilodr\'an-Casas, C\'esar and Ouala, Said and Farchi, Alban and Liu, Che and Tandeo, Pierre and Fablet, Ronan and Lucor, Didier and Iooss, Bertrand and Brajard, Julien and Xiao, Dunhui and Janjic, Tijana and Ding, Weiping and Guo, Yike and Carrassi, Alberto and Bocquet, Marc and Arcucci, Rossella},
  title   = {Machine learning with data assimilation and uncertainty quantification for dynamical systems: a review},
  journal = {IEEE/CAA Journal of Automatica Sinica},
  volume  = {10},
  number  = {6},
  pages   = {1361--1387},
  year    = {2023},
}

@book{evensen2022fundamentals,
  author    = {Evensen, Geir and Vossepoel, Femke C. and van Leeuwen, Peter Jan},
  title     = {Data Assimilation Fundamentals: {A} Unified Formulation of the State and Parameter Estimation Problem},
  publisher = {Springer Nature},
  year      = {2022},
}

@book{kalnay2003,
  author    = {Kalnay, Eugenia},
  title     = {Atmospheric Modeling, Data Assimilation and Predictability},
  publisher = {Cambridge University Press},
  year      = {2003},
}

@article{carrassi2018,
  author  = {Carrassi, Alberto and Bocquet, Marc and Bertino, Laurent and Evensen, Geir},
  title   = {Data assimilation in the geosciences: an overview of methods, issues, and perspectives},
  journal = {WIREs Climate Change},
  volume  = {9},
  number  = {5},
  pages   = {e535},
  year    = {2018},
}

@article{burgers1998,
  author  = {Burgers, Gerrit and van Leeuwen, Peter Jan and Evensen, Geir},
  title   = {Analysis scheme in the ensemble {K}alman filter},
  journal = {Monthly Weather Review},
  volume  = {126},
  number  = {6},
  pages   = {1719--1724},
  year    = {1998},
}

@article{iglesias2013,
  author  = {Iglesias, Marco A. and Law, Kody J. H. and Stuart, Andrew M.},
  title   = {Ensemble {K}alman methods for inverse problems},
  journal = {Inverse Problems},
  volume  = {29},
  number  = {4},
  pages   = {045001},
  year    = {2013},
}

@article{zhang2022ensemble,
  author  = {Zhang, Xin-Lei and Xiao, Heng and Luo, Xiaodong and He, Guowei},
  title   = {Ensemble {K}alman method for learning turbulence models from indirect observation data},
  journal = {Journal of Fluid Mechanics},
  volume  = {949},
  pages   = {A26},
  year    = {2022},
}

@inproceedings{west2025variational,
  author    = {West, John F. and Buchta, David A. and Zaki, Tamer A.},
  title     = {Variational data assimilation in shock tube flows},
  booktitle = {AIAA SCITECH 2025 Forum},
  year      = {2025},
}

@article{edoh2026,
  author  = {Edoh, Ayaboe K. and West, John F. and Houba, Thomas and Munipalli, Ramakanth and Harvazinski, Matthew E. and Kang, Wei},
  title   = {Data assimilation of compressible flows with discontinuities: evaluating algorithms on {S}od's shock tube},
  journal = {International Journal for Numerical Methods in Fluids},
  volume  = {98},
  number  = {4},
  pages   = {644--672},
  year    = {2026},
}

@article{hansen2025detonation,
  author  = {Hansen, Jon J. and Brouzet, Davy and Ihme, Matthias},
  title   = {Data assimilation of state-indirect observations for constraining detonation wave dynamics},
  journal = {Proceedings of the Combustion Institute},
  volume  = {41},
  pages   = {105812},
  year    = {2025},
}

@article{subrahmanya2025,
  author  = {Subrahmanya, Amit N. and Sandu, Adrian},
  title   = {Feature preserving data assimilation via feature alignment},
  journal = {Computer Methods in Applied Mechanics and Engineering},
  volume  = {447},
  pages   = {118345},
  year    = {2025},
}

@article{mousavi2025,
  author  = {Mousavi, Ali and Jones, Brandon A. and Eldredge, Jeff D.},
  title   = {Sequential estimation of aerodynamic flows with a reduced latent space},
  journal = {arXiv preprint arXiv:2509.03795},
  year    = {2025},
}

@article{bach2025,
  author  = {Bach, Eviatar and Baptista, Ricardo and Calvello, Edoardo and Chen, Yifan and Stuart, Andrew M.},
  title   = {Learning enhanced ensemble filters},
  journal = {Journal of Computational Physics},
  pages   = {114550},
  year    = {2025},
}

@article{zhou2024bieqno,
  author  = {Zhou, Xu-Hui and Liu, Zhuo-Ran and Xiao, Heng and Wang, Qiqi},
  title   = {{BI-EqNO}: Generalized approximate {B}ayesian inference with an equivariant neural operator framework},
  journal = {arXiv preprint arXiv:2410.16420},
  year    = {2024},
}

@article{zhou2026ensemble,
  title={Ensemble generative filtering for sequential data assimilation in dynamical systems},
  author={Zhou, Xu-Hui and Han, Jiequn},
  journal={arXiv preprint arXiv:2609.14078},
  year={2026}
}

@article{glaubitz2026regularity,
  title={Regularity-informed data assimilation: {A} hierarchical {B}ayesian approach to ensemble {K}alman filtering for hyperbolic conservation laws},
  author={Glaubitz, Jan and Sharp, Daniel and {Le Provost}, Mathieu and Marzouk, Youssef M.},
  journal={arXiv preprint arXiv:2608.14840},
  year={2026}
}

@article{stordal2011adaptive,
  author  = {Stordal, Andreas S. and Karlsen, Hans A. and N{\ae}vdal, Geir and Skaug, Hans J. and Vall{\`e}s, Brice},
  title   = {Bridging the ensemble {K}alman filter and particle filters: the adaptive {G}aussian mixture filter},
  journal = {Computational Geosciences},
  volume  = {15},
  number  = {2},
  pages   = {293--305},
  year    = {2011},
}

@article{smith2007cluster,
  author  = {Smith, Kevin W.},
  title   = {Cluster ensemble {K}alman filter},
  journal = {Tellus A: Dynamic Meteorology and Oceanography},
  volume  = {59},
  number  = {5},
  pages   = {749--757},
  year    = {2007},
}

@article{sondergaard2013parti,
  author  = {Sondergaard, Thomas and Lermusiaux, Pierre F. J.},
  title   = {Data assimilation with {G}aussian mixture models using the dynamically orthogonal field equations. {P}art {I}: Theory and scheme},
  journal = {Monthly Weather Review},
  volume  = {141},
  number  = {6},
  pages   = {1737--1760},
  year    = {2013},
}

@article{sondergaard2013partii,
  author  = {Sondergaard, Thomas and Lermusiaux, Pierre F. J.},
  title   = {Data assimilation with {G}aussian mixture models using the dynamically orthogonal field equations. {P}art {II}: Applications},
  journal = {Monthly Weather Review},
  volume  = {141},
  number  = {6},
  pages   = {1761--1785},
  year    = {2013},
}

@article{lee2020model,
  author  = {Lee, Kookjin and Carlberg, Kevin T.},
  title   = {Model reduction of dynamical systems on nonlinear manifolds using deep convolutional autoencoders},
  journal = {Journal of Computational Physics},
  volume  = {404},
  pages   = {108973},
  year    = {2020},
}

@article{champion2019data,
  author  = {Champion, Kathleen and Lusch, Bethany and Kutz, J. Nathan and Brunton, Steven L.},
  title   = {Data-driven discovery of coordinates and governing equations},
  journal = {Proceedings of the National Academy of Sciences},
  volume  = {116},
  number  = {45},
  pages   = {22445--22451},
  year    = {2019},
}

@article{lusch2018deep,
  author  = {Lusch, Bethany and Kutz, J. Nathan and Brunton, Steven L.},
  title   = {Deep learning for universal linear embeddings of nonlinear dynamics},
  journal = {Nature Communications},
  volume  = {9},
  pages   = {4950},
  year    = {2018},
}

@article{maulik2021reduced,
  author  = {Maulik, Romit and Lusch, Bethany and Balaprakash, Prasanna},
  title   = {Reduced-order modeling of advection-dominated systems with recurrent neural networks and convolutional autoencoders},
  journal = {Physics of Fluids},
  volume  = {33},
  number  = {3},
  pages   = {037106},
  year    = {2021},
}

@article{fresca2021comprehensive,
  author  = {Fresca, Stefania and Ded\`{e}, Luca and Manzoni, Andrea},
  title   = {A comprehensive deep learning-based approach to reduced order modeling of nonlinear time-dependent parametrized {PDE}s},
  journal = {Journal of Scientific Computing},
  volume  = {87},
  pages   = {61},
  year    = {2021},
}

@article{vlachas2022multiscale,
  author  = {Vlachas, Pantelis R. and Arampatzis, Georgios and Uhler, Caroline and Koumoutsakos, Petros},
  title   = {Multiscale simulations of complex systems by learning their effective dynamics},
  journal = {Nature Machine Intelligence},
  volume  = {4},
  pages   = {359--366},
  year    = {2022},
}

@article{wiewel2019latent,
  author  = {Wiewel, Steffen and Becher, Moritz and Thuerey, Nils},
  title   = {Latent space physics: {T}owards learning the temporal evolution of fluid flow},
  journal = {Computer Graphics Forum},
  volume  = {38},
  number  = {2},
  pages   = {71--82},
  year    = {2019},
}

@article{hotelling1936relations,
  author  = {Hotelling, Harold},
  title   = {Relations between two sets of variates},
  journal = {Biometrika},
  volume  = {28},
  number  = {3-4},
  pages   = {321--377},
  year    = {1936},
}

@article{bjorck1973numerical,
  author  = {Bj\"{o}rck, {\AA}ke and Golub, Gene H.},
  title   = {Numerical methods for computing angles between linear subspaces},
  journal = {Mathematics of Computation},
  volume  = {27},
  number  = {123},
  pages   = {579--594},
  year    = {1973},
}

@article{udaykumar2001sharp,
  author  = {Udaykumar, H. S. and Mittal, R. and Rampunggoon, P. and Khanna, A.},
  title   = {A sharp interface {C}artesian grid method for simulating flows with complex moving boundaries},
  journal = {Journal of Computational Physics},
  volume  = {174},
  number  = {1},
  pages   = {345--380},
  year    = {2001},
}



\end{document}